\documentclass[12pt,preprint]{aastex}

\shorttitle{A FUV Survey of M\,80} 
\shortauthors{Dieball et al.}

\newcommand{\HST}{{\it HST}}

\begin{document}

\title{A Far-UV Survey of M\,80: X-ray Source Counterparts, Strange
  Blue Stragglers and the Recovery of Nova T\,Sco.\footnote{Based
    on observations with the NASA/ESA Hubble Space Telescope, obtained
    at the Space Telescope Science Institute, which is operated by the
    Association of Universities for Research in Astronomy, Inc. under
    NASA contract No. NAS5-26555.}}

\author{Andrea Dieball}
\affil{Department of Physics and Astronomy, University Southampton,
  SO17 1BJ, UK}
  
\author{Knox S. Long}
\affil{Space Telescope Science Institute, Baltimore, MD 21218}

\author{Christian Knigge, Grace S. Thomson}
\affil{Department of Physics and Astronomy, University Southampton,
  SO17 1BJ, UK}

\and

\author{David R. Zurek}
\affil{Department of Astrophysics, American Museum of Natural History,
New York, NY 10024}

\begin{abstract}

Using the ACS on \HST, we have surveyed the $FUV$ and $NUV$
populations in the core region of M\,80. The CMD reveals large numbers
of blue and extreme horizontal branch stars and blue stragglers, as
well as $\approx60$ objects lying in the region of the CMD where
accreting and detached white dwarf binaries are expected. Overall, the
blue straggler stars are the most centrally concentrated population,
with their radial distribution suggesting a typical blue straggler
mass of about $1.2~M_\odot$. However, counterintuitively, the faint
blue stragglers are significantly more centrally concentrated than the
bright ones and a Kolmogorov-Smirnov test suggest only a 3.5\%
probability that both faint and bright blue stragglers are drawn from
the same distribution. This may suggest that (some) blue stragglers
get a kick during their formation. We have also been able to identify
the majority of the known X-ray sources in the core with $FUV$ bright
stars. One of these $FUV$ sources is a likely dwarf nova that was in
eruption at the time of the $FUV$ observations. This object is located
at a position consistent with Nova 1860 AD, or T\,Scorpii. Based on
its position, X-ray and UV characteristics, this system is almost
certainly the source of the nova explosion. The radial distribution of
the X-ray sources and of the cataclysmic variable candidates in our
sample suggest masses $> 1 M_\odot$.

\end{abstract}


\section{Introduction}

Far-UV ($FUV$) observations are an ideal tool to study the exotic
stellar populations that reside in globular clusters (GCs), such as
blue stragglers (BSs), white dwarfs (WDs), cataclysmic variables (CVs,
binaries containing an accreting WD), low-mass X-ray binaries (LMXBs,
binaries containing an accreting neutron star or black hole), blue and
extreme horizontal branch (BHB and EHB) stars, and blue hook (BHk)
stars. Identifying these exotica in visible light can be extremely
difficult, partly due to the severe crowding of optical images
(especially of the cores of GCs) which are dominated by main sequence
(MS) stars and red giants (RGs), and partly because most exotica are
optically faint. However, all of these exotic populations tend to be
hotter than other cluster members and emit much of their radiation in
the $FUV$. ``Ordinary'' cluster stars (MS stars and RGs) are cooler
and considerably fainter at wavelengths less than 2000 \AA. As a
result, crowding is generally not a problem in the $FUV$. Thus, deep
$FUV$-imaging at high spatial resolution with $HST$ is an excellent
way to detect and study exotic stellar species.

So far, deep $FUV$ studies have been carried out for only three
clusters: 47\,Tuc (Knigge et al.~2002, 2003, 2008), NGC\,2808 (Brown
et al. 2001, Dieball et al. 2005a, Servillat et al. 2008), and M\,15
(Dieball et al.~2005b, 2007). In 47\,Tuc, Knigge et al.~(2002) found
$FUV$ counterparts for the four {\it Chandra} CV candidates (Grindlay
et al. 2001) known at that time within the $FUV$ field of view. All of
these were found to be variable $FUV$ excess sources and three of them
were later spectroscopically confirmed as CVs (Knigge et
al.\ 2008)\footnote{The fourth likely CV was located outside the field
  of view of the spectroscopic observations.}. In NGC\,2808, Brown et
al.~(2001) used $FUV$ and $NUV$ imaging to uncover a population of
sub-luminous hot horizontal branch (HB) stars, the BHk stars, and
suggested that this population is the result of a late helium-core
flash on the WD cooling curve. Dieball et al. (2005a) re-analyzed the
$HST$ data on NGC\,2808 and found numerous BSs, CV candidates and hot
young WDs. Servillat et al.~(2008) found 8 $FUV$ counterparts to the
Chandra X-ray sources in the core of NGC\,2808; two of those are close
matches and confirm their CV nature. In M\,15, Dieball et al.~(2005b)
found the $FUV$ counterpart of the LMXB M\,15~X-2 (White \& Angelini
2001), and clearly detected an orbital period of 22.6 minutes, thus
confirming M\,15~X-2 as an ultracompact X-ray binary (UCXB), only the
third in a GC at that time. Since then, Zurek et al.~(2009) have
confirmed the UCXB status of another X-ray source in the GC NGC\,1851,
also based on FUV observations. Dieball et al.~(2007) constructed a
deep $FUV - NUV$ color magnitude diagram (CMD) for M\,15, which
revealed large numbers of CV and WD candidates, a well defined BS and
HB sequence, and 41 variable $FUV$ sources, amongst them RR Lyrae,
Cepheids, SX Phoenicis stars, CVs, and the well known LMXB AC\,211.

Here we present the results of $FUV$ and $NUV$ imaging observations
with the Advanced Camera for Surveys (ACS) with the $Hubble~Space
~Telescope ~ (HST)$ of the GC M\,80. This cluster is one of the
densest in the Galaxy and has a metallicity of $[\rm{Fe/H}] = -1.7$
dex (Brocato et al.\ 1998, Alcaino et al.\ 1998, Cavallo et
al.\ 2004), a distance of 10 kpc, and a reddening of $E_{B-V} = 0.18$
mag (Harris 1996). Despite being a very dense and compact cluster
($r_{core} = 9\arcsec$ corresponding to 0.44 pc at 10 kpc,
$r_{halfmass} = 39\arcsec$ corresponding to 1.89 pc, Harris 1996),
M\,80 is not thought to be a core-collapsed cluster. Ferraro et
al. (1999, 2003) found a large and centrally concentrated population
of BSs, and suggested that M\,80 is in a state in which core-collapse
is delayed by the production of an extraordinarily large population of
collisional BSs. Only a few variable sources are known in M\,80 (Wehlau
et al.\ 1990, Clement \& Walker 1991, Clement et al. 2001). Based on
the periods of the six RR Lyrae known in this cluster, M\,80 is
classified as Oosterhoff type II (Oosterhoff 1939).

M\,80 is also famous for its historic classical nova T\,Scorpii,
which was discovered in 1860 by Auwers when it outshone the entire
cluster (Luther 1860, Pogson 1860). Shara \& Drissen (1995) found a
very blue star $5\arcsec$ from the cluster center and within
$1\arcsec$ of their estimated position for the nova, and thus
identified it as the (now quiescent) counterpart of the nova. Our
observations, discussed in Section \ref{xray}, suggest a different
candidate. Apart from the nova, two new erupting dwarf novae (DNe)
were identified by Shara \& Drissen (1995). Based on a 50 ksec of
Chandra observations, Heinke et al. (2003) found 19 X-ray sources
within the cluster's halfmass radius to a limiting
$L_{\rm{0.5-2.5\,keV}}\approx7\times10^{30}\ \rm{erg}\ \rm{s}^{-1}$.
They suggested that two of those were quiescent LMXBs and five others
were CVs based on their X-ray hardness ratio, and that the brightest
source detected might be the X-ray counterpart to the classical nova
T\,Sco.
 
Our report is structured as follows: In Sect.~\ref{data}, we describe
the observations and the data reduction. In Sect.~\ref{cmd}, the
analysis of the $FUV - NUV$ CMD is presented. In Sect.~\ref{xray},
we describe our comparison of X-ray to $FUV$ locations and our
identification of the quiescent nova with an object which appears to
have been undergoing a dwarf nova outburst at the time of our $FUV$
observations. In Sect.~\ref{radial}, we present the radial
distribution of the various populations and compare them to the X-ray
source distribution. Finally, we summarize our results in
Sect.~\ref{summary}.

\section{Observations and the Creation of the Catalog}
\label{data}

The observations of the core of M\,80 were carried out with the ACS
onboard {\it HST} using the $FUV$ F165lp filter in the Solar Blind
Channel (SBC) and the $NUV$ F250W filter in the High Resolution
Channel (HRC) and were made at a single pointing position. The SBC has
a field of view of $35\arcsec \times 31\arcsec$ with a pixel size of
$0\farcs034 \times 0\farcs030$, whereas the HRC field of view is
slightly smaller with $29\arcsec \times 26\arcsec$ and a spatial
resolution of $0\farcs028 \times 0\farcs025$ pixels. Thus, the
observations cover only the central portions (approximately 1.5 core
radii if we adopt a core radius of $9\arcsec$, Harris 1996) of
M\,80. The $FUV$ observation was carried out during four consecutive
orbits in September 2004. To facilitate searches for time variability,
the $FUV$ observation (dataset j8y501) comprised 32 individual
exposures of durations ranging from 310-323 seconds. The total
exposure time was 10232 sec. The $NUV$ observation (dataset j8y504)
comprised a single orbit in October 2004, and resulted in a total
exposure of 2384 sec, split into 8 individual exposures of 298
sec. Dithers were not utilized to simplify searches for time
variability.
 
Beginning with the data products delivered by STScI, we created master
images of the $FUV$ and $NUV$ data using {\tt multidrizzle} running
under {\tt PyRAF}. The {\tt multidrizzle} routines correct the field
distortion that exists in the individual flatfielded images delivered
as part of the standard data products and combine them into master
images for the $FUV$ and $NUV$. The combined and geometrically
corrected output master images have a pixel scale of $0.025\arcsec /
{\rm pixel}$ and are normalized to 1 sec exposure time.

\begin{figure}
\includegraphics[width=16cm]{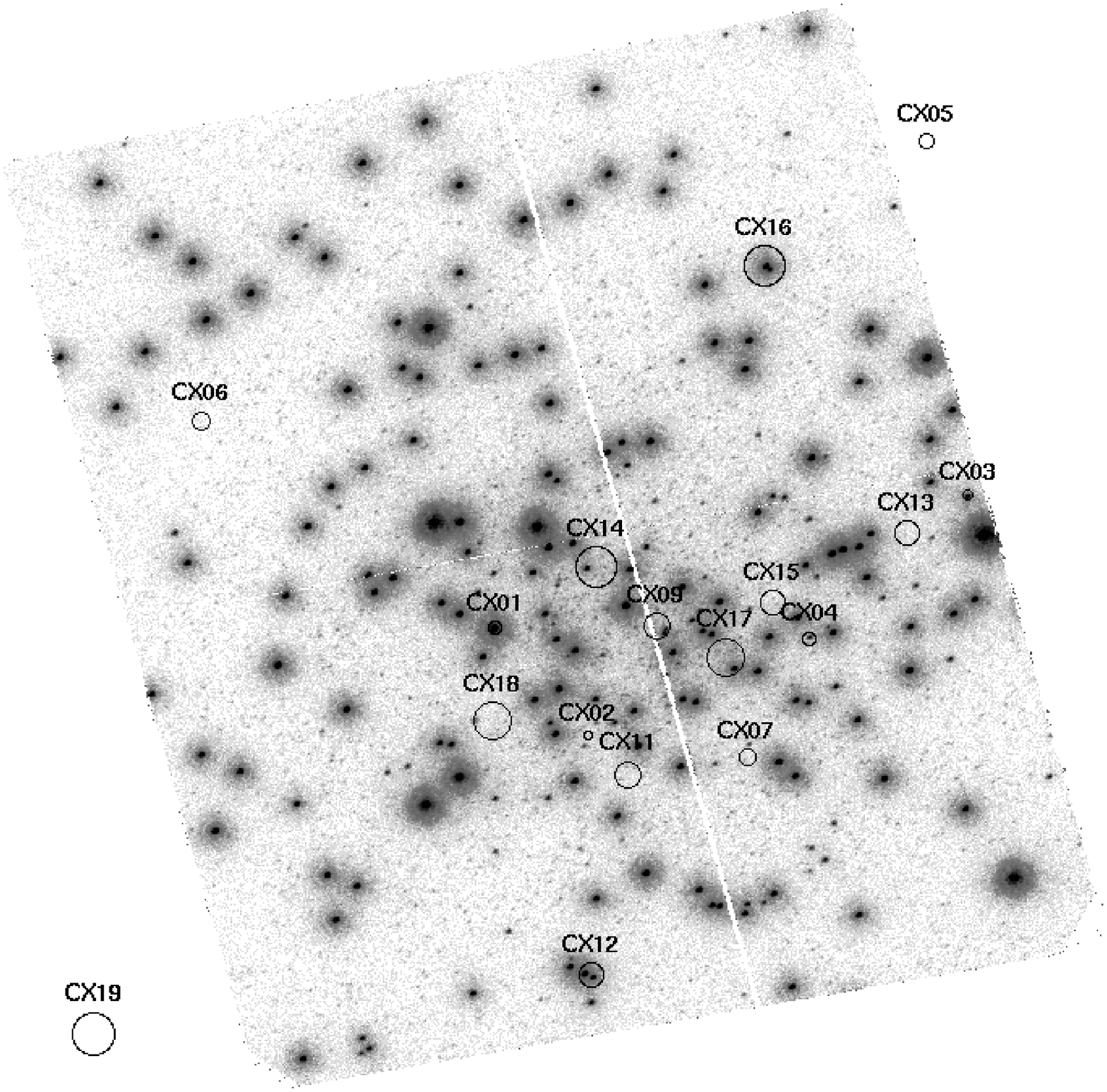}
\caption{ Combined and geometrically corrected master image of all
  $FUV$ SBC/F165LP exposures taken from M\,80's core region. North is
  up and east to the left. The field of view is $35\arcsec \times
  31\arcsec$. The image is displayed on a logarithmic intensity scale
  in order to bring out the fainter sources. The positions of the
  X-ray sources found by Heinke et al. (2003) are marked with their
  3$\sigma$ error circles. \label{fig_FUV}}
\end{figure}

\begin{figure}
\includegraphics[width=16cm]{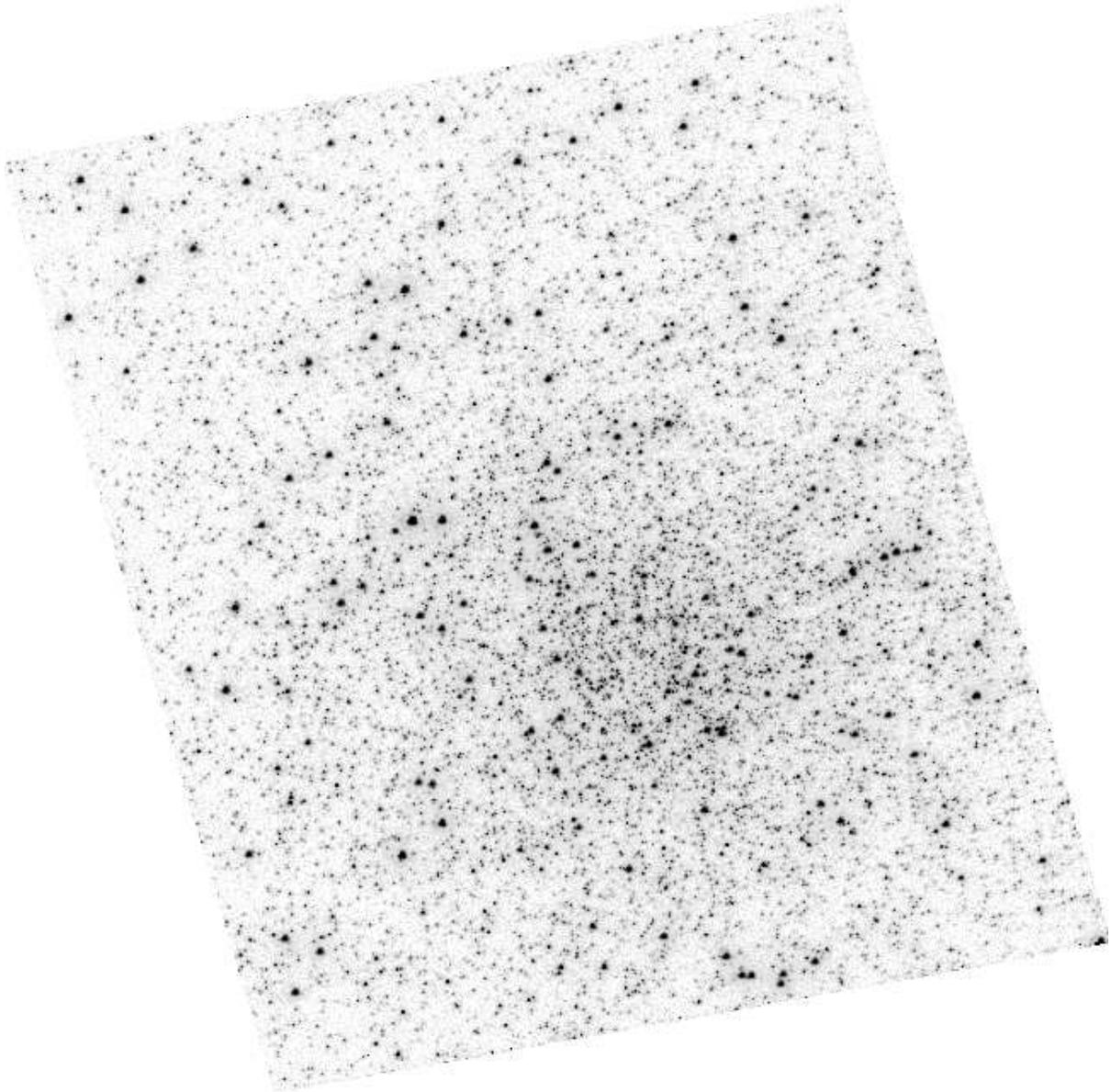}
\caption{ Same as Fig.~\ref{fig_FUV}, but for the
  $NUV$ HRC/F250W. Note that the HRC field of view is somewhat smaller than the
  SBC field of view with $29\arcsec \times 25\arcsec$. \label{fig_NUV}}    
\end{figure}

The $FUV$ and $NUV$ master images are shown in Fig.~\ref{fig_FUV} and
\ref{fig_NUV}. As expected, the $FUV$ image is considerably less
crowded than the $NUV$ image. Both images show significant
concentrations of sources towards the cluster core.

\subsection{$FUV$ and $NUV$ Source Detection}
\label{detection}

We used {\tt daofind} (Stetson 1991) running under {\tt
  IRAF}\footnote{{\tt IRAF} (Image Reduction and Analysis Facility) is
  distributed by the National Astronomy and Optical Observatories,
  which are operated by AURA, Inc., under cooperative agreement with
  the National Science Foundation.} to create initial source lists for
the $FUV$ and $NUV$ master images. We then checked and updated these
lists, adding a few faint stars that were missed by {\tt daofind} and
removing obvious false detections (e.g. multiple detections of very
bright sources, noise peaks near image edges, etc.). For a detailed
description of the source finding procedure see Dieball et
al.\ (2007). The resulting final catalogs contained 3168 $FUV$ and
9875 $NUV$ sources.

\subsection{Matching $FUV$, $NUV$ and Optical Sources}
\label{matching}

In order to match the $FUV$ and $NUV$ catalogs, we created a reference
list containing the pixel coordinates of 92 stars that are clearly
visible and well within the fields of both images. We used the {\tt
  geomap} and {\tt geoxytran} task running under {\tt IRAF} to
determine the geometrical transformation between the two catalogs. We
allowed for x and y shifts, rotation and scale changes in the
coordinate transformation. The residual errors in the transformation
were quite small, less than 0.2 pixels ($<$ 10 mas) (RMS) for the 92
stars.

The $FUV$ field is slightly larger than the $NUV$ field and 2574 $FUV$
sources are located within the $NUV$ field of view. After some
testing, we adopted a maximum matching tolerance of 2.5 pixels between
the $FUV$ and $NUV$ source positions, resulting in 2345 matches (91\%
of the possible $FUV$ sources). Following the procedure described by
Knigge et al. (2002), which is based both on the number of sources
which are matched and those which are not matched, we can expect 45
($\approx 1.9\%$) false matches among these 2345 pairs (but note that
this estimate does not account for the increased source concentration
towards the core).
 
We also used the Piotto et al. (2002) catalog of M\,80 to search for
optical counterparts to our $FUV$ sources. The optical data were
obtained using the WFPC2 in 1996, with the PC centered on the cluster
center. As a first step, our $FUV$ image coordinate system had to be
transformed to the PC image system. For that purpose, we used 31 HB
stars as reference objects that could be easily identified in both the
PC F555W and the SBC F165LP master image. (Note that five out of these
31 stars are not located within the somewhat smaller field of view of
the HRC F250W exposures.) We allowed for a maximum matching tolerance
of 1.1 PC pixels (corresponding to $\approx2$ pixels on our $FUV$ or
$NUV$ master images) and found a total of 1418 optical matches to the
$FUV$ sources; out of these 1268 are inside the $NUV$ field of
view. We can expect $\approx 40$ to be false matches.

\subsection{Improving the Absolute Astrometry}
\label{positions}

Even though {\tt multidrizzle} corrects the field distortion of our
images, it does not improve their absolute astrometric accuracy. The
world coordinate system (WCS) of the images provided with the standard
data products is based on the original guide star catalog (GSC1),
whose absolute positions are often only accurate to $1\arcsec -
2\arcsec$. This makes matching to external (e.g. X-ray) catalogs
difficult.

The usual way to improve the astrometry in HST images is to locate one
or more stars in an image whose positions are accurately known in a
Tycho-based system, and to update the astrometric solution of the
image. However, because the core of M\,80 is so crowded and because
the SBC covers such a small region of the sky, we were unable to find
appropriate stars in the master images. We therefore adopted a
bootstrap approach beginning with an ACS Wide Field Camera (WFC) F435W
image of M\,80 (namely HST\_10573\_03\_ACS\_WFC\_F435W\_sci.fits)
obtained from the Hubble Legacy Archive. The WFC has a field of view
of $202\arcsec \times 202\arcsec$, and the image covers not only the
core of M\,80, but also regions around the core where the density of
stars is considerably lower. As a result, we were able to locate 16
stars from the Second US Naval Observatory CCD Astrograph Catalog
(UCAC2, Zacharias et al.~2004) in the WFC image. The UCAC2 catalog is
tied to the Tycho system and has an absolute astrometric error of
$\approx 70$ milliarcseconds (mas) for stars brighter than R of 16
mag. Based on the positions of the UCAC2 stars in this field, we
updated the astrometric solution for the ACS WFC, updating the
boresight for the ACS WFC image by approximately $1\farcs2$. The RMS
error between positions of the stars in our UCAC2 sample, which were
scattered fairly uniformly around the core of M\,80 in the WFC image,
was approximately $0\farcs2$.

We then located 16 (non-saturated) stars that could be easily
identified in both the $NUV$ HRC and the WFC image, and used their
positions to remove the offset and distortion between the $NUV$ HRC
and the WFC image. In doing this, we allowed for image offsets,
rotation, and linear scale changes. The same procedure and the same
stars were used to correct the WCS of the $FUV$ SBC image. The RMS
error between the positions of the stars in the $NUV$ image and the
WFC image following this correction was 12 mas. Unless otherwise
noted, all of the positional information we discuss here has been
obtained from master images with the corrected WCSs. We conservatively
estimate our overall error to be less than $0\farcs2$.

A catalog listing all our $FUV$ objects is available in the online
version of ApJ. For reference, we list only 20 entries in
Table~\ref{tab_catalog}.   

\begin{rotate}
\begin{deluxetable}{ccccccccccrccl}
\tablecaption{Catalog of all sources in our $FUV$ field of view. The
  first column is the line number in the catalog, followed by the
  $FUV$ id number in col.~2. Cols.~3 to 6 give the source position in
  RA and DEC and image pixel coordinates, cols.~7 to 10 give the $FUV$
  and $NUV$ magnitudes and corresponding photometric errors as derived
  from {\tt daophot}. Col.~11 gives the id number of the optical
  counterpart taken from Piotto et al.~(2002), followed by the optical
  magnitudes in cols.~12 and 13. The final column 14 include the
  source type according to its position in the $FUV - NUV$ and optical
  CMD and further comments. Only 20 entries are listed. \label{tab_catalog}}
\tabletypesize{\scriptsize}
\tablewidth{0pt} 
\tabcolsep0.1cm 
\tablehead
{1 & 2 & 3 & 4 & 5 & 6 & 7 & 8 & 9 & 10 & 11 & 12 & 13 & 14\\
ID$_{cat}$ & ID$_{FUV}$ & RA & DEC & $x_{FUV}$ & $y_{FUV}$& $FUV$ &$\Delta FUV$&$NUV$&$\Delta NUV$&ID$_{Piotto}$&B&V & Comments \\ 
 & & [hh:mm:ss] & [deg:mm:ss] & [pixels] & [pixels] & [mag] & [mag] & [mag] & [mag] & & [mag] & [mag] & \\ } 
\startdata 
100 & 702  & 16:17:01.593 & -22:58:43.05 & 1384.857 & 329.680 & 23.771 & 0.226 & *      & *     & *    & *      & *      & no NUV, outside PC\\ 
101 & 1972 & 16:17:01.598 & -22:58:34.98 & 1381.465 & 652.124 & 24.322 & 0.306 & *      & *     & *    & *      & *      & outside HRC \\ 
102 & 1895 & 16:17:01.599 & -22:58:35.45 & 1380.791 & 633.057 & 23.169 & 0.142 & *      & *     & 2548 & 20.553 & 19.674 & outside HRC \\ 
103 & 2818 & 16:17:01.603 & -22:58:29.23 & 1378.636 & 881.567 & 17.614 & 0.011 & *      & *     & *    & *      & *      & outside HRC \\ 
104 & 382  & 16:17:01.603 & -22:58:45.19 & 1379.270 & 244.316 & 23.155 & 0.139 & *      & *     & *    & *      & *      & outside HRC, outside PC\\ 
105 & 2932 & 16:17:01.607 & -22:58:28.50 & 1376.075 & 910.782 & 24.201 & 0.328 & *      & *     & *    & *      & *      & outside HRC \\ 
106 & 1400 & 16:17:01.607 & -22:58:38.54 & 1376.704 & 509.832 & 23.023 & 0.129 & 21.123 & 0.030 & 2157 & 20.765 & 20.018 & MS \\ 
107 & 1811 & 16:17:01.608 & -22:58:35.97 & 1376.083 & 612.429 & 24.293 & 0.274 & 22.038 & 0.135 & *    & *      & *      & MS/RG clump \\ 
108 & 1125 & 16:17:01.608 & -22:58:40.34 & 1376.593 & 437.772 & 16.385 & 0.006 & 17.662 & 0.003 & *    & *      & *      & EHB, outside PC\\ 
109 & 430  & 16:17:01.608 & -22:58:44.88 & 1376.571 & 256.750 & 22.817 & 0.118 & *      & *     & *    & *      & *      & outside HRC, outside PC\\ 
110 & 2334 & 16:17:01.610 & -22:58:32.60 & 1374.507 & 747.079 & 22.896 & 0.234 & *      & *     & *    & *      & *      & outside HRC \\ 
111 & 1461 & 16:17:01.612 & -22:58:38.14 & 1374.040 & 525.815 & 24.754 & 0.333 & 22.035 & 0.057 & *    & *      & *      & MS/RG clump \\ 
112 & 1639 & 16:17:01.612 & -22:58:37.07 & 1374.187 & 568.424 & 24.108 & 0.204 & *      & *     & *    & *      & *      & no NUV \\ 
113 & 3025 & 16:17:01.613 & -22:58:27.88 & 1372.837 & 935.394 & 23.976 & 0.227 & *      & *     & 3213 & 21.048 & 20.070 & outside HRC \\ 
114 & 2083 & 16:17:01.615 & -22:58:34.26 & 1372.089 & 680.526 & 24.004 & 0.250 & *      & *     & 2668 & 21.693 & 20.476 & outside HRC \\ 
115 & 887  & 16:17:01.615 & -22:58:41.80 & 1372.850 & 379.650 & 22.505 & 0.108 & 18.622 & 0.006 & *    & *      & *      & MS/RG clump, outside PC\\ 
116 & 1261 & 16:17:01.620 & -22:58:39.48 & 1369.875 & 472.383 & 22.852 & 0.132 & 20.557 & 0.018 & *    & *      & *      & MS/RG clump \\ 
117 & 1591 & 16:17:01.621 & -22:58:37.34 & 1369.040 & 557.550 & 23.209 & 0.142 & 21.172 & 0.024 & *    & *      & *      & MS/RG clump \\ 
118 & 354  & 16:17:01.623 & -22:58:45.35 & 1368.608 & 238.054 & 23.411 & 0.164 & *      & *     & *    & *      & *      & outside HRC, outside PC\\ 
119 & 1767 & 16:17:01.625 & -22:58:36.24 & 1366.857 & 601.537 & 23.346 & 0.155 & 20.354 & 0.014 & *    & *      & *      & MS/RG clump \\ 
120 & 947  & 16:17:01.626 & -22:58:41.42 & 1366.732 & 394.774 & 24.767 & 0.413 & *      & *     & *    & *      & *      & no NUV, outside PC\\ 
\enddata
\end{deluxetable}
\end{rotate}

\subsection{The Cluster Center}
\label{centre}

Several estimates for the cluster center exist in the literature and
have been compiled in Table~\ref{tabcenter}. The Shawl \& White
(1986)\footnote{Their estimate for the cluster center was adopted by
  Djorgovski \& Meylan (1993) and Harris (1996).}  estimate is based
on smoothed scans of ESO/SRC photographic plates; the Ferraro et
al. (1999) estimate is the average of the stellar coordinates measured
in an HST/WFPC2/PC image of the cluster core; the Shara \& Drissen
(1995) estimate is based on smoothed isophotes created from an
HST/WFPC2/PC image of the core. The Ferraro et al. (1999) and Shara \&
Drissen (1995) coordinates are both in the Guide Star Catalogue
system, while the Shawl \& White (1986) coordinates are based on
stellar positions in the SAO catalog.

Since the Tycho system to which we have tied our data is superior to
both the SAO and the GSC systems, we have redetermined the cluster
center, using our own observations. The $NUV$ data set, in particular,
is eminently suitable for this purpose, since it contains a
sufficiently large number of stars, yet is not seriously affected by
crowding. We thus estimate the position of the cluster center by
maximizing the number of $NUV$ sources contained in a circular region
of radius $r_{lim}$ when the center of this region is varied. For our
final estimate, we adopted $r_{lim}$ = 150 pixels ($= 3\farcs75$), but
other reasonable choices yield consistent results. The uncertainties
on our center coordinates were estimated by a simple bootstrapping
method. More specifically, we created 1000 fake $NUV$ catalogs by
sampling with replacement from the actual catalog. We then estimated
the center for each of the bootstrapped mock catalogs in the same way
as for the real data. The standard deviation of the mock center
estimates is then adopted as the error. In this way, we determined the
cluster center at $x_{F250W} = 789 \pm 13$ and $y_{F250W} = 533 \pm
18$, which corresponds to $\alpha = 16^{h}17^{m}02.432^{s} \pm
0\farcs325$, $\delta = -22^{\circ}58\arcmin34\farcs62 \pm 0\farcs45$
in our Tycho-based WCS (see Sect.~\ref{positions}).

\begin{deluxetable}{ccl}
\tablecaption{Estimates of the cluster center.\label{tabcenter}} 
\tablewidth{0pt}
\tablehead{RA                & DEC                        & reference}
\startdata
$16^{h}17^{m}02.432^{s}$ & $-22^{\circ}58\arcmin34\farcs62$ & this paper\\
$16^{h}17^{m}02.29^{s}$  & $-22^{\circ}58\arcmin32\farcs38$ & Ferraro et al. (1999)\\
$16^{h}17^{m}02.48^{s}$  & $-22^{\circ}58\arcmin33\farcs8$  & Shara \& Drissen (1995)\\
$16^{h}17^{m}02.51^{s}$  & $-22^{\circ}58\arcmin30.4\arcsec$ & Shawl\& White (1986)\\
\enddata
\end{deluxetable}

\subsection{Aperture Photometry}
\label{photometry}

Photometry was performed on the combined and geometrically corrected
$FUV$ and $NUV$ master images using {\tt daophot} (Stetson 1991)
running under {\tt IRAF}. Because of the high density of objects in
the core region of M\,80 (especially in the $NUV$ image), we used a
small aperture radius of 3 pixels, and a small sky annulus of 5 -- 7
pixels. We also allowed for a Gaussian recentering of the input
coordinates. For the $FUV$ data, we determined corrections for the
finite aperture size and the source flux contained in the sky annulus
from a few bright and isolated stars in the master image. The
procedure is described in more detail in Dieball et al.~(2007). For
the $NUV$ data, we use the same method to determine the sky
correction, but adopted the aperture corrections published by Sirianni
et al. (2005). Note that our $FUV$ aperture correction only reaches
out to 60 (SBC) pixels, since larger apertures are invariably affected
by bright neighbors. However, our curve of growth analysis for the
bright, isolated stars in the $FUV$ image suggests that the
additional aperture correction from 60 pixels to infinity is small. By
contrast, Sirianni et al.~(2005) give aperture corrections to a
maximum aperture of $0\farcs5$ and suggest an additional correction to
infinity of 0.132 mag that should be added to the derived STMAG in
HRC/F250W. Since we do apply this infinite radius correction to
the $NUV$ data, there could be a slight systematic red bias in our
$FUV$-$NUV$ colors. All magnitudes are given on the STMAG system,
where 
\begin{eqnarray}
\nonumber
\rm{STMAG} = -2.5 \times log_{10} (count rate \times PHOTFLAM \times
apcorr \times skycorr) + ZPT + addcorr.
\end{eqnarray}  
The correction and conversion factors we used to convert count rates
into fluxes and STMAGs are listed in Table~\ref{corr}.   

We estimate the overall completeness limits in our catalog to $FUV
\approx 23$ mag and $NUV \approx 21.5$ mag. However, since the
detection of sources in the broad PSF wings of the bright sources in
the $FUV$ is extremely difficult, the completeness limit is not strictly
uniform and considerably lower near bright $FUV$ sources. 

\begin{deluxetable}{cccccc}
\tablecaption{\label{corr} Conversion and correction factors. The
  conversion factor PHOTFLAM (column 2) is needed to convert count
  rates into fluxes. The zero point ZPT is given in column 3, followed
  by the encircled energy fraction $ee = 1/apcorr$ within a 3 pixel
  aperture radius (column 4), and the sky correction from 5 -- 7 to 50
  -- 60 pixel sky annulus (column 5). The final column gives the
  additional magnitude correction from a $0\farcs5$ aperture to
  infinity, as suggested by Sirianni et al.~(2005). See the text for
  details.}  
\tablewidth{0pt} 
\tablehead{camera/filter & PHOTFLAM & ZPT & ee & skycorr & addcorr\\ 
& [$\rm{erg\,cm}^{2} \AA^{-1} \rm{counts}^{-1}$] & [mag] & & & [mag]} 
\startdata 
SBC/F165LP & 1.3596913E-16 & 21.1 & 0.47$\pm$0.02 & 1.029$\pm$0.005 & - \\  
HRC/F250W & 4.7564122E-18 & 21.1 & 0.655$\pm$0.006 & 1.017$\pm$0.003 & -0.132$\pm$0.002\\ 
\enddata
\end{deluxetable}

\section{The $FUV-NUV$ CMD}
\label{cmd}

\begin{figure}
\centerline{
\includegraphics[width=17cm,height=14.5cm]{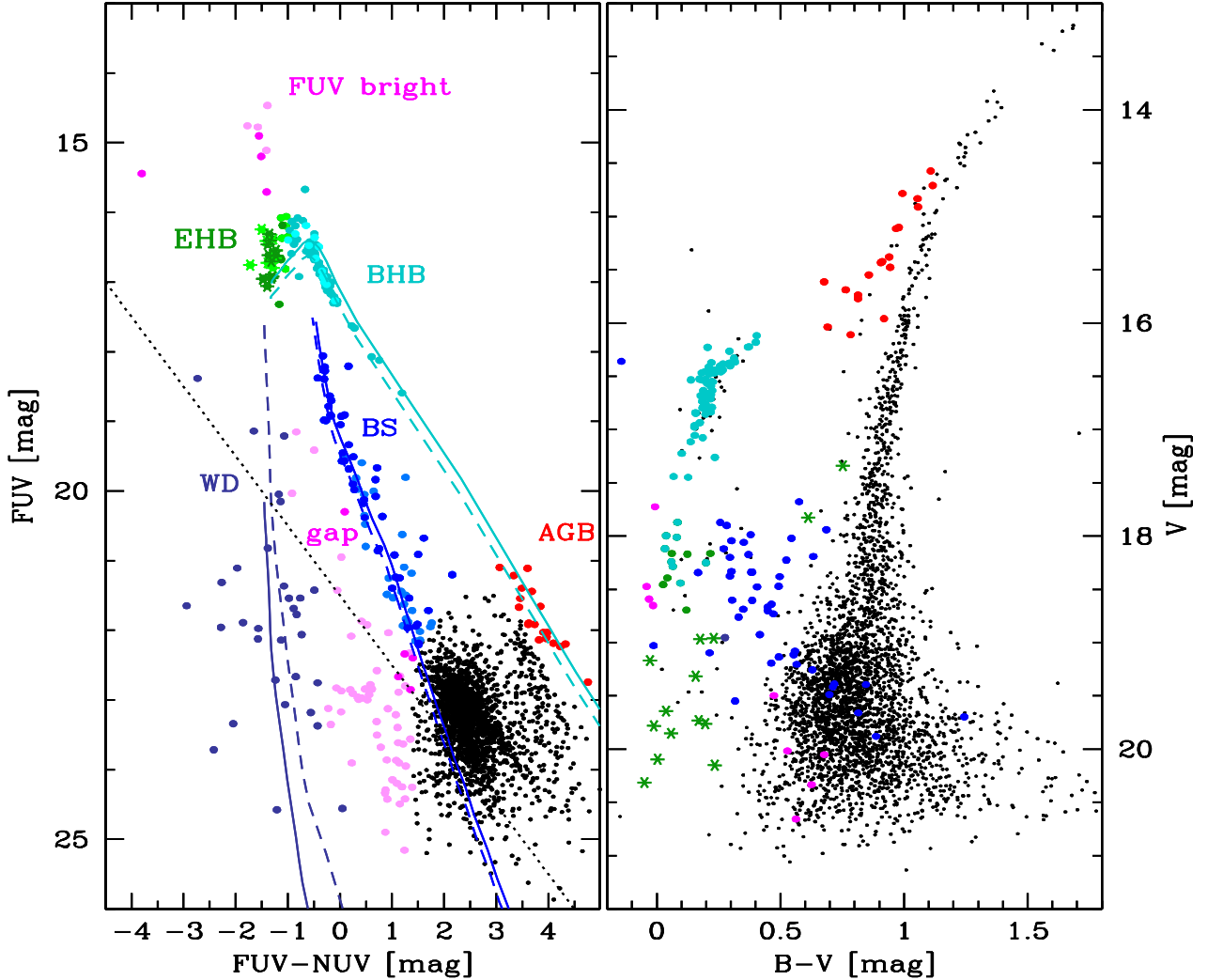}
}
\caption{\label{cmda} Left panel: $FUV-NUV$ CMD of the core
  region of M\,80. For orientation purposes, we include a theoretical
  WD and He WD cooling sequence (violet lines), a zero-age main
  sequence (ZAMS, blue line), and a zero-age HB track (ZAHB, cyan
  line). BHB stars are plotted in cyan, EHB stars in green, BSs in
  blue, gap sources (which include CV candidates) in magenta, WD
  candidates in violet, and AGB stars in red. The $FUV$ bright sources,
  which are likely AGB manch{\'e} stars, are plotted in magenta. The
  remaining sources are MS stars and RGs. $FUV$ sources with optical
  counterparts are plotted in a darker shade of the same color (except
  for MS stars and RGs). Right panel: Optical CMD of M\,80. The
  data were taken from Piotto et al. (2002), only the PC data are
  plotted. The counterparts to the $FUV$ sources are plotted in the
  same color as in the left diagram. See the text for details.}
\end{figure}

\begin{figure}
\centerline{
\includegraphics[width=17cm,height=14.5cm]{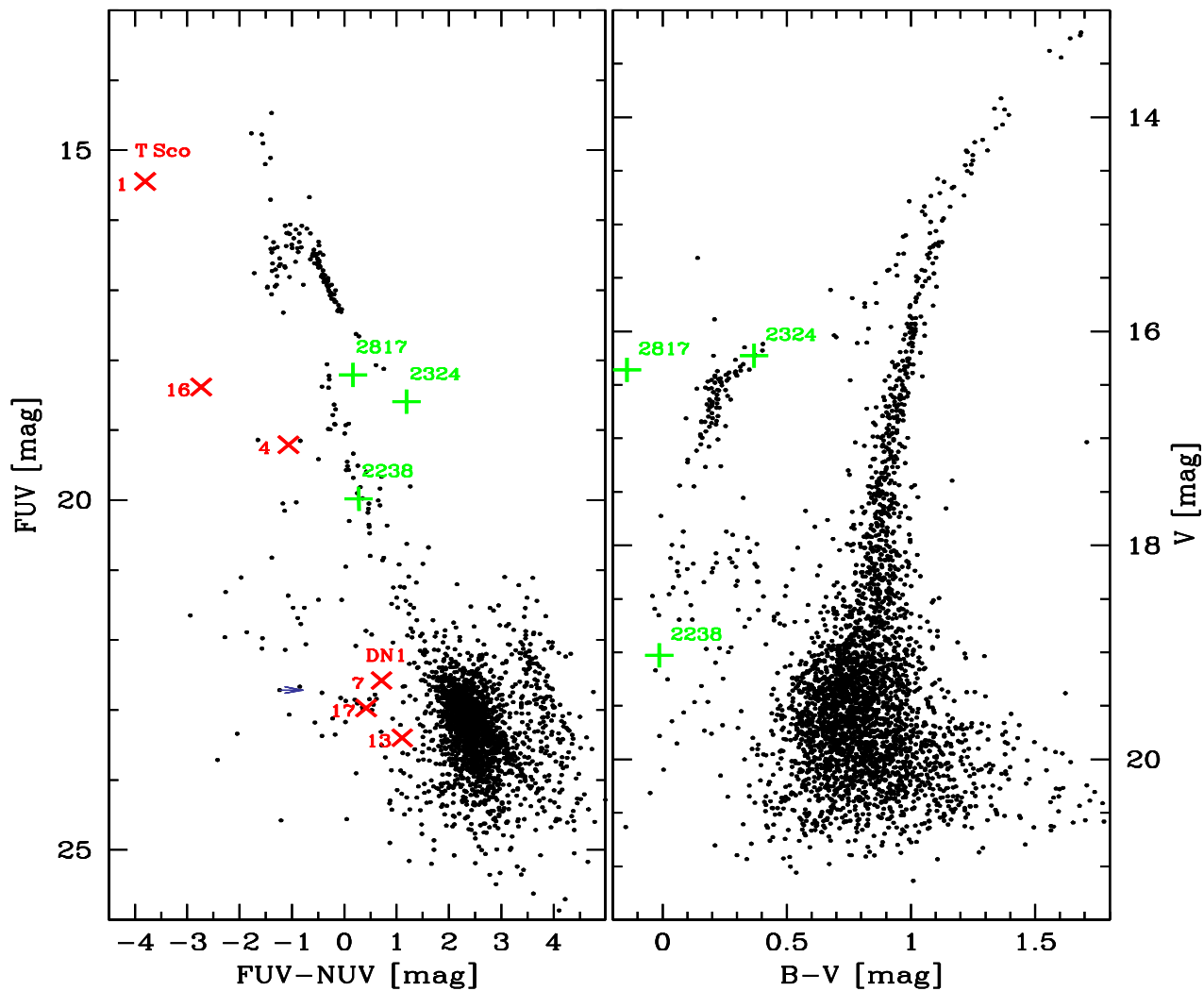}
}
\caption{\label{cmdb} Same as Fig.~\ref{cmda}, but with the variable
  sources marked with green crosses and their $FUV$ id. The most
  likely counterparts to the X-ray sources are marked with tilted red
  crosses and their X-ray source id. The WD candidate that is located
  on the rim of the repeller wire shadow in the $NUV$ image, and that
  might thus actually be $NUV$ brighter and redder, is marked with a
  small violet arrow. See the text for details.}
\end{figure}

The $FUV - NUV$ CMD of the core region of M\,80 is shown in
Figs.~\ref{cmda} and \ref{cmdb} (left diagrams). For orientation
purposes, we include a theoretical zero-age MS (ZAMS, plotted in blue
in Fig.~\ref{cmda}) and a WD (solid violet line in Fig.~\ref{cmda})
and Helium white dwarf (He WD) cooling sequence (dashed violet
line). For all our synthetic tracks we adopted a distance of 10 kpc, a
reddening of $E_{B-V}=0.18$ mag and a cluster metallicity of $[Fe/H]
\simeq -1.7$ (Harris 1996). For details on these tracks, see Dieball
et al. (2005a). We also plot a zero-age HB (ZAHB, cyan line) which was
constructed based on the $\alpha$ enhanced BaSTI ZAHB model for
[Fe/H]=1.62 dex and a mass loss parameter $\eta = 0.4$ (e.g. Cordier
et al. 2007). We then used {\tt synphot} within {\tt IRAF} to
calculate the corresponding $FUV$ and $NUV$ STMAGs. As can be seen in
Fig.~\ref{cmda}, the ZAHB and ZAMS appear to be somewhat brighter
and/or redder. Increasing the distance makes the ZAHB and ZAMS
fainter, whereas decreasing the reddening makes the tracks brighter
and bluer. We found that using a somewhat larger distance of 11.5 kpc
and slightly smaller reddening of $E_{B-V}=0.17$ mag for the ZAHB and
ZAMS, plotted as dashed tracks in Fig.~\ref{cmda}, gives a better fit
to our data. However, we point out that the synthetic tracks are
plotted for orientation purposes, and - given the difficulties in
calibrating the $UV$ data - we do not aim to (re)determine distance
and reddening from our $FUV - NUV$ CMD.

The $FUV - NUV$ CMD clearly contains various stellar populations,
including WD candidates (violet data points in Fig.~\ref{cmda}, left
diagram), BSs (blue data points), HB stars (plotted in green and
cyan), and AGB stars (red) that are located at the faint and red end
of the ZAHB. The CMD also contains a group of objects between the WD
cooling sequence and the ZAMS. This is the expected location of WD --
MS binaries. We call these objects ``gap sources'' because the CMD
alone does not allow us to distinguish between mass exchanging
binaries (CVs) and non-interacting WD binaries (see, e.g. Dieball et
al. 2007). The remaining sources (black data points) in the $FUV -
NUV$ CMD are MS stars and RGs. The MS turnoff is at $FUV \approx 22.5$
mag and can be recognized by the sudden increase in source numbers
along the ZAMS around that magnitude. Our CMD reaches approximately
2.5 mags fainter than the MS turnoff in the $FUV$.

The optical CMD is shown in the right panel of Figs.~\ref{cmda} and
\ref{cmdb}. The optical counterparts to $FUV$ objects are marked with
the same color as in the $FUV - NUV$ CMD. The location of the stellar
populations in the optical CMD agrees well with what we expect based
on the $FUV - NUV$ CMD, e.g. the counterparts to the $FUV$ BHB stars
are on the optical BHB as well, and also the location of the EHB, AGB,
and BS stars agrees in both the $FUV - NUV$ and the optical CMD.

In the following subsections, we will discuss the HB, BS, WD and gap
source populations in more detail. The selection of stars belonging to
the various populations is based on the $FUV - NUV$ CMD, but the
numbers we give for the populations should not be taken as exact,
since the various zones in the $FUV$ CMD overlap and the
discrimination between them can be difficult.
 
\subsection{The Horizontal Branch in the $FUV$}
\label{hb}

Our $FUV - NUV$ CMD contains a significant population of both EHB and
BHB stars located along the bright ($FUV < 19$ mag) part of the
ZAMS. We define stars to be EHB stars if they have colors at least as
blue as the ZAHB at $T_{eff} \approx 20,000$ K (e.g. Momany et
al. 2004), corresponding to $FUV - NUV = -1.0$ mag in our CMD. As
can be seen in Fig.~\ref{cmda}, the optical CMD shows a large gap
along the vertical BHB/EHB tail, as was already noted by Ferraro et
al. (1998). This gap appears in the $FUV - NUV$ CMD as
well and occurs around the ``knee'' of the ZAMS at $FUV - NUV \approx
-0.7$ mag. In both the $FUV - NUV$ and the optical CMD, another,
optically fainter and $FUV$ bluer gap is visible. In the $FUV - NUV$
CMD, the bluer gap occurs in the EHB part of the ZAHB sequence
approximately at $FUV - NUV \approx -1.2$ mag, corresponding to $T_{eff}
\approx 26,000$ K in our model. Sources with (photometrically) hotter
$T_{eff}$ are plotted as stars in both CMDs. We refer to these sources
as EHB2 stars. EHB stars redder than the optical faint/$FUV$ blue gap
are denoted EHB1 stars (see Sect.~\ref{radial}). As can be seen, the
bluer EHB2 stars agree very well with the optically fainter EHB
stars. Two of the optical counterparts to the EHB2 stars are located
in the BS region close to the RGB and might be mismatches.

Fig.~\ref{hb} (top panel) shows a zoom on the Horizontal Branch in our
$FUV - NUV$ CMD. For comparison, we plot the $FUV - V$ CMD in the
bottom panel and mark the location of the gaps visible in our data,
and of the four gaps described in Ferraro et al. (1998) according to
their temperatures along the BaSTI ZAHB. All gaps are marked with a
solid arrow in both the $FUV - V$ and $FUV - NUV$ CMDs and are denoted
as $_{\rm{D}}$ if identified in our data, and $_{\rm{F}}$ if referring to
the gaps discussed in Ferraro et al. 1998. None of the gaps visible in
our CMD appear at the same temperatures as suggested by Ferraro et
al. (1998), instead we found that the gaps appear to be somewhat
shifted. Ferraro et al. (1998) suggested a temperature of 9500 K for
their gap G0$_{\rm{F}}$. We see a gap close to this temperature
position only in the $FUV-V$ data at 10000 K, our gap G0$_{\rm{D}}$,
but this gap is not visible in our $FUV-NUV$ CMD. G1$_{\rm{F}}$, at
11000 K in Ferraro et al. (1998), cannot be identified in our $FUV-V$
CMD, but we caution that the low number of BHB stars between
$-0.5>FUV-V>-1.5$ prevent a secure detection. Our $FUV-NUV$ CMD does
not indicate a gap in that area. As already noted, a large gap is
visible around $FUV-NUV \approx -0.7$ and $FUV - V \approx -1.4$ that
we denote G2$_{\rm{D}}$. According to the temperatures along the ZAHB,
this G2$_{\rm{D}}$ is in between Ferraro's G2$_{\rm{F}}$ and
G3$_{\rm{F}}$ gap. Also, in our data G3 appears at somewhat bluer
colours and higher temperatures. Table~\ref{hbtab} gives an overview
of the temperatures assigned to the gaps in Ferraro et al. (1998,
their Fig. 4) and this work, all colours refer to the corresponding
temperatures based on the BaSTI ZAHBs. Please note that Ferraro et
al. (1998) shifted their M\,80 $FUV - V$ CMD to match that of M\,13,
and they also use a different $FUV$ filter (the WFPC2 F160BW). The
differences between the exact temperature location of the gaps is
likely due to differences in the $FUV$ filter\footnote{Ferraro et
  al. (1998) used WFPC2 F160BW filter that has a pivot wavelength 1522
  \AA\ and a bandwidth 449\AA, whereas our $FUV$ data were obtained
  with ACS, SBC, F165LP which has a pivot wavelength of 1758~\AA\ and
  a bandwidth of 86 \AA.}, the HB models used (Ferraro et al. used the
Dorman et al. 1993 models, whereas we use the newer BaSTI models), the
parameters assumed for the HB model (distance, reddening), and also
the calibration of the data. As can be seen, the BaSTI ZAHB fit the
$FUV - V$ data somewhat better than the $FUV - NUV$ data, suggesting
that the $NUV$ aperture correction might be underestimated, rendering
the $FUV - NUV$ color too blue. Keeping this in mind, the temperatures
assigned to the gaps (G0 and G2) match relatively well, except for the
bluest gap G3 which we found at higher $T_{eff}$, more comparable to
the G3 gap in NGC\,2808 (see Ferraro et al. 1998, their table 2).

\begin{deluxetable}{lccccc}
\tablecaption{\label{hbtab} Gap colors and temperatures in our $FUV -
  NUV$ and $FUV - V$ CMDs (cols. 2 -- 4). Col. 5 denotes the color
  corresponding to the temperature associated with the Ferraro et
  al. (1998) gaps.}  
\tablewidth{0pt} 
\tabcolsep0.15cm 
\tablehead{ & \multicolumn{3}{c}{this paper} &
  \multicolumn{2}{c}{Ferraro et al. 1998}\\
 & FUV-NUV & FUV-V & $T_{eff}$ & FUV-V & $T_{eff}$}
\startdata
 G0 & --     &  0.059 & 10000 & 0.329  &  9500\\
 G1 & --     &  --    & --    & -0.389 & 11000\\
 G2 & -0.721 & -1.405 & 14500 & -0.743 & 12000\\
 G3 & -1.178 & -2.814 & 25500 & -2.019 & 18000\\ 
\enddata
\end{deluxetable}

Based on the WFPC2 data presented by Ferraro et al. (1998), Momany et
al. (2004) suggested that M\,80 might contain BHk stars. BHk stars are
as blue as the EHB stars, but $FUV$ fainter (see Brown et
al. 2001). If these stars exist in M\,80, our observations show they
are very rare. Our $FUV - NUV$ CMD shows only one star that is fainter
than the hot end of the ZAHB. This could either be a somewhat fainter
but ``normal'' EHB star, or a BHk candidate. Unfortunately, we did not
find an optical counterpart for this source. Definite BHk stars have
so far only been found in the most massive GCs, although the number of
GCs surveyed does not allow one to conclude that lower mass GCs
are incapable of producing BHk stars (see Dieball et al.\ 2009).

\subsection{Blue Stragglers}
\label{bs}

Fig.~\ref{cmda} shows a well defined trail of stars above the MS
turnoff and around the ZAMS, with a few sources located slightly to
the red of the ZAMS. This is the expected location of BSs in a CMD if
they are produced via the collision or coalescence of two or more
lower-mass MS stars. As they are more massive than the MS stars, we
expect them to be slightly evolved. We have marked 75 sources as
likely BSs; 47 of these have optical counterparts which agree very
well with the expected location in the optical CMD. Some of the
optical counterparts are fainter than the optical MS turnoff. These
are likely BSs with progenitors less massive than the MS turnoff
mass. BSs in GCs are thought to be formed dynamical via stellar
collisions and/or from evolution of primordial binaries. It is still
subject to discussion which is the dominant formation mechanism, if
there is one. Recent studies have found no correlation of BSs numbers
(or frequencies) with the collision rate, arguing against dynamical
formation as the dominating channel (Piotto et al. 2004, Leigh et
al. 2007), but on the other hand the radial distribution of BSs seems
to be bimodal in many clusters, with a strong central peak, indicating
that dynamics play an important role in the formation of BSs in the
cores of these clusters (e.g. Dalessandro et al. 2008, Mapelli et
al. 2006, Lanzoni et al. 2007, Ferraro et al. 2004).

Ferraro et al. (1999) found an unusual large and centrally
concentrated fraction of BSs in M\,80 (305 BSs in their WFPC2
dataset). They suggest that these BSs are collisionally formed, and
that M\,80 is currently in a transient dynamical state where core
collapse is delayed via stellar interactions which led to the
formation of the large number of BSs (but also see Knigge et
al.\ 2009, who suggest that most BSs - even in the core of GCs - are
descended from binary stars, although they do not rule out stellar
dynamics as a key factor in the formation and evolution of the parent
binaries). More recently, Ferraro et al. (2003) compared six GCs
(M\,3, M\,80, M\,10, M\,13, M\,92, and NGC\,288) and found that M\,80
has the largest and most concentrated population of BSs. We compare
our $FUV$ M\,80 data to our $FUV$ M\,15 data, which covered a similar
field of view as well, and do not find a remarkable excess in BSs. In
both M\,80 and M\,15 we find the same number of BSs (75), but the two
clusters are different in the sense that M\,15 is even more massive,
more concentrated, and more metal-poor compared to M\,80. Scaling with
the field size at the distance of the cluster, the ratio of BS
numbers in M\,80 and M\,15 is only slightly above unity and not
significantly different from the ratio obtained for WDs, for example
(see Table~\ref{gc}). This, at first glance, argues against an
anomalous enhancement of BS numbers in M\,80, at least compared to
M\,15. On the other hand, if the BS specific frequencies as defined in
Ferraro et al. (1999), $F^{BS}_{HB} = N_{BS}/N_{HB}$, are considered,
M\,15 shows a slightly lower BS specific frequency.  This agrees with
Piotto et al. (2004) who found an anticorrelation of BS frequency and
cluster mass, and a (mild) tendency of increasing BS frequency with
decreasing central density. However, allowing for a Poisson error on
the number of stars found within the clusters, the difference between
the $F^{BS}_{HB}$ is $0.08\pm0.12$. Thus, although the difference in
$F^{BS}_{HB}$ between M\,80 and M\,15 seems to reflect the trend
discussed in Piotto et al. (2004), it is still not statistically
significant.

\begin{deluxetable}{lcccccccccccccc}
\tablecaption{\label{gc} Number of HB and gap sources, WD and BS
  candidates (cols. 2 -- 5) detected in M\,15 and M\,80, and the
  number of sources per pc$^{2}$ (cols. 6 -- 8). Col. 9 gives the BS
  specific frequency, cols. 10 -- 13 give the cluster distance,
  metallicity, and the logarithmic core and halfmass relaxation time lg(tc),
  and lg(th) (Harris 1996), col. 14 the cluster total mass (Gnedin et
  al. 2002), and col. 15 the the central density lg($\rho_{c}$).}
\tablewidth{0pt} 
\tabcolsep0.11cm 
\tablehead{ 
name &HB &gap&WD&BS&$\frac{\rm{gap}}{\rm{pc}^{2}}$&$\frac{\rm{WD}}{\rm{pc}^{2}}$&$\frac{\rm{BS}}{\rm{pc}^{2}}$&$F^{BS}_{HB}$&distance&[Fe/H]&lg(tc)&lg(th)&M$_{tot}$&lg($\rho_{c}$)\\ 
     &   &   &  &  &            &          &           &            &[kpc]   &      &      &      &         & } 
\startdata 
M\,15&133&48 &34&75& 26         & 18.5     & 40.7      & 0.564      & 10.3   &-2.26 & 7.02 & 9.35 &1.19     &5.38\\ 
M\,80&117&59 &31&75& 34         & 17.3     & 43.2      & 0.641      & 10.0   &-1.77 & 7.73 & 8.86 &0.50     &4.76\\ 
\enddata
\end{deluxetable}

\subsection{White Dwarfs}
\label{wd}

We find $\approx 30$ sources that are located close to the WD and He
WD sequences in Fig.~\ref{cmda}. Most of these are likely to be WDs,
although a few could be CVs or detached WD -- MS binaries (as we will
argue in Sect.~\ref{xray}). The number of expected WDs in our field of
view can be estimated from the number of HB stars and the relative
lifetimes of stars on the HB and the cooling timescale for WDs. In
total, we have 117 HB sources (30 EHB stars, 80 BHB stars and 7 $FUV$
bright sources that are likely AGB manch{\'e} stars). In making this
comparison, we can only consider WD candidates above the completeness
limits in {\em both} $FUV$ and $NUV$ images. If we therefore restrict
our WD sample to a limiting magnitude of $FUV \approx 22$ mag
(corresponding to $T_{eff} \approx 24,000$ K and a cooling age of
$2\times10^{7}$ yrs), we find 24 WD candidates in our CMD; which
compares very well to the 23 WDs that are expected on the basis of the
HB numbers. This suggests that most, if not all, of our candidates are
indeed WDs. Note that one of our WDs has an unexpectedly bright
optical counterpart with $V \approx 19$ mag. However, the $NUV$
counterpart is located at the rim of the repeller wire shadow in the
$NUV$ image, and might thus actually be brighter. In this case, this
source would be redder, which would shift it into the BS region in our
$FUV - NUV$ CMD. This source is marked with an arrow in
Fig.~\ref{cmdb} (left diagram).

\subsection{Gap Sources - CVs and other WD Binaries}
\label{cv}

A number of sources can be seen in the ``gap'' between the WD cooling
sequences and the ZAMS in Fig.~\ref{cmda}. As mentioned earlier, we
cannot distinguish between the CV candidates and the detached WD -- MS
binaries, so we call these objects the ``gap sources''. How many CV
candidates can we expect amongst the $\approx 60$ gap sources?
Detailed theoretical work was done on 47\,Tuc (di\,Stefano \&
Rappaport 1994, Shara \& Hurley 2006, Ivanova et al. 2006),
investigating various dynamical and primordial formation channels for
CVs and predicting a few 100 CVs in this cluster. For the sake of
simplicity, we adopt di Stefano \& Rappaport's (1994) prediction of
190 active CVs in 47\,Tuc, and scale this number with the capture rate
(e.g. Heinke et al. 2003) to M\,80. This yields $\approx 100$ CVs that
can be expected in M\,80. According to di\,Stefano \& Rappaport
(1994), approximately half of the captures should take place inside
the cluster core. Given that our detection limit corresponds to a
white dwarf temperature of $T_{eff} \approx 24,000$ K (see
Sect.~\ref{wd}), we will only be able to detect relatively bright,
long-period CVs above the period gap (Townsley \& Bildsten, 2003,
their Figs.~1 and 2). About twenty of these long-period CVs should
exist in 47\,Tuc (Di\,Stefano \& Rappaport, 1994, their Fig.~3 and
Table~5). Ivanova et al. (2006) suggested that 35 œôòó- 40 CVs should be
detectable in the core of 47 Tuc. Scaling these numbers to M\,80, we
can expect approximately 10 to 20 such sources in the core of
M\,80. Note that the $NUV$ field of view covers $\approx 1.5$ times
the core radius of M\,80. Thus, the number of objects we find in the
gap region is consistent with the number of predicted CVs. Five of our
gap sources have optical counterparts, all of which lie bluewards of
the MS and below the optical MS turnoff. These systems might have
relatively massive MS companions that dominate the optical light. On
the other hand, all five of these sources are located close to the
ZAMS in the $FUV$ CMD, making them BS candidates as well.

\subsection{Variable Sources}

As noted earlier, only few variable sources are known in M\,80 (Wehlau
et al.\ 1990, Clement \& Walker 1991, Clement et al. 2001), and indeed
the region where RR Lyrae stars are expected is unpopulated in our
$FUV - NUV$ CMD. Our $FUV$ observations cover four consecutive HST
orbits, comprising 32 single images, and we have used these data to
search for variable sources. We found three sources that exhibit
convincing evidence of variability, namely source no. 2238, 2324, and
2817. These sources are flagged as variable in
Table~\ref{tab_catalog}, and are marked in Fig.~\ref{cmdb}. Source
no. 2238 shows short-term variability ($\approx55$ min) and is likely
a SX\,Phoenicis star. Source no. 2324 shows long-term variability and
might be a RR Lyrae or a Cepheid. Source no. 2817 is a peculiar object
that shows very strong variability. A more detailed study on the
variable sources in M\,80, including their lightcurves, will be
presented in a forthcoming paper (Thomson et al., in preparation).

\section{Identification of X-ray Sources \label{xray}}

As noted earlier, M\,80 was observed with $Chandra$ by \cite{heinke}
who identified 19 discrete sources within the half-mass radius of the
cluster. All but four of these -- CX05, CX08, CX10, and CX19 -- are in
the field of the $FUV$ image. Since the majority of the X-ray sources
are expected to be CVs, and all are thought to be binaries, identifying
their counterparts at longer wavelengths is important, and our $UV$
images and source catalogs provide us with an excellent opportunity to
accomplish this task. 

Heinke et al.\ (2003) referenced the X-ray source positions in M\,80
to a bright star (HD146457) in the Tycho catalog that was roughly
$4\arcmin$ from the core of M\,80, and allowed for an absolute
position error of $2\arcsec$. This error is quite large, given the
crowding of the core of M\,80, so we have attempted to find a more
accurate way to register the X-ray sources to our fields. HD146457 is
not in the ACS image we used to establish an accurate (Tycho-based)
WCS for our UV images of the core of M\,80. We therefore compared the
positions of the 52 X-ray sources identified by Heinke et al.\ (2003)
outside of the core, which presumably are mostly foreground stars and
background quasars, to the ACS WFC 435W field. Eight of these sources
are located within the region covered by the WFC image, and two of
those, J161658.3-225838 and 161659.8-225931, were near relatively
bright stars. The offset between the Heinke positions and these two
stars was approximately $0\farcs13$ in $\alpha$ and $1\farcs17$ in
$\delta$, well within Heinke's estimated error.\footnote{\cite{heinke}
  did not attempt a similar comparison. They corrected the X-ray
  positions using HD16457 and used the \cite{shara95} positions for
  the nova which were based on GSC1.}

After correcting the Heinke et al.\ (2003) X-ray source positions by
these values, we compared the X-ray sources with the $FUV$ image of
M\,80. It was immediately apparent that a number of the X-ray sources
had counterparts with the brighter sources in the field.  Of the
fifteen X-ray sources within our $FUV$ field of view, six lie within
$1\arcsec$ of a bright $FUV$ source ($FUV < 22$ mag) that has no
optical counterpart. We then applied a final shift of $0\farcs1$ to
optimally align the three X-ray sources with the most definite $FUV$
counterparts (CX01, CX03 and CX04). (Note that these are 3 of the 4
brightest X-ray sources, and that CX02, the one not associated with a
bright $FUV$ source, has a spectrum which suggests it is a quiescent
LMXB). The positions of the X-ray sources on the $FUV$ image after
these corrections are shown in Fig. \ref{fig_FUV}, where the circles
represent the 3$\sigma$ statistical uncertainty in X-ray position as
determined by Heinke et al. (2003). As explained in more detail below,
there are 6 X-ray sources with bright $FUV$ counterparts, whose
identifications we consider secure. The final RMS offset between the
X-ray and $FUV$ positions of these 6 sources is only $0\farcs14$ after
alignment.

We then compared the positions of all X-ray sources to those of
objects in our $FUV$ catalog, using the 3$\sigma$ statistical
uncertainty in X-ray position for each source. Table~\ref{tab_xray}
summarizes the results of this comparison. For some sources,
especially those which are faint in X-rays, and hence have larger
error circles, multiple sources lie within the 3$\sigma$ radius. In
these cases, we have listed all of the possible $FUV$ counterparts in
order of increasing distance from the nominal X-ray position. The four
Chandra sources which were outside the $FUV$ field of view appear in
the table, but only to indicate their improved positions.

\begin{rotate}
\begin{deluxetable}{cccccrccccrccl}
\tablecaption{Chandra X-ray Source Comparison. The X-ray source id is
  given in the first column, followed by our revised positions for the
  X-ray sources in cols.~2 and 3, the 3$\sigma$ statistical
  uncertainty in col.~4, the angular distance from the nominal
  position to the $FUV$ object in col.~5, the following cols.~6 to 14
  are as in Table~\ref{tab_catalog}. \label{tab_xray}} 
\tabletypesize{\scriptsize}
\tablewidth{0pt}
\tablehead
{1 & 2 & 3 & 4 & 5 & 6 & 7 & 8 & 9 & 10 & 11 & 12 & 13 & 14\\
ID$_{X}$ &  RA     &  DEC         & $3 \sigma$  & Offset      & ID$_{FUV}$& $FUV$& $\Delta FUV$ & $NUV$  & $\Delta NUV$ & ID$_{Piotto}$& B    & V      & Comments\\
     & [hh:mm:ss]   & [deg:mm:ss]  & [$\arcsec$] & [$\arcsec$] &        & [mag]  & [mag]        & [mag]  & [mag]        &           & [mag]  & [mag]  & \\ }
\startdata
CX01 & 16:17:02.817 & -22:58:33.92 &  0.22       &  0.02       &  2129  & 15.444 & 0.005        & 19.247 &      0.008   &  *        & *      & *      &  FUVbright \\ 
CX02 & 16:17:02.580 & -22:58:37.73 &  0.13       &  0.08       &  1523  & 23.736 & 0.229        & 21.247 &      0.029   &  *        & *      & *      &  MS/RG \\ 
CX03 & 16:17:01.600 & -22:58:29.20 &  0.18       &  0.05       &  2818  & 17.614 & 0.011        & *      &      *       &  *        & *      & *      &  outside HRC \\ 
CX04 & 16:17:02.008 & -22:58:34.28 &  0.23       &  0.04       &  2082  & 19.209 & 0.022        & 20.277 &      0.024   &  *        & *      & *      &  WD \\ 
~    & ~            & ~            &  ~          &  0.22       &  4790  & 22.589 & 0.134        & 18.748 &      0.007   &  2190     & 16.289 & 15.198 &  RG \\ 
CX05 & 16:17:01.711 & -22:58:16.59 &  *          &  *          &  *     & *      & *            & *      &      *       &  *        & *      & *      &  * \\ 
CX06 & 16:17:03.573 & -22:58:26.55 &  0.29       &  0.21       &  3221  & 23.656 & 0.181        & 21.152 &      0.031   &  *        & *      & *      &  MS/RG clump\\ 
~    & ~            & ~            &  0.29       &  0.25       &  3181  & 23.448 & 0.162        & 21.210 &      0.024   &  *        & *      & *      &  MS/RG clump \\ 
CX07 & 16:17:02.169 & -22:58:38.52 &  0.27       &  0.12       &  1387  & 22.578 & 0.120        & 21.869 &      0.055   &  *        & *      & *      &  gap \\ 
~    & ~            & ~            &  ~          &  0.25       &  4850  & 22.926 & 0.159        & 20.666 &      0.020   &  *        & *      & *      &  MS/RG clump  \\ 
CX08 & 16:17:01.118 & -22:58:30.58 &  *          &  *          &  *     & *      & *            & *      &      *       &  *        & *      & *      &  * \\ 
CX09 & 16:17:02.404 & -22:58:33.85 &  0.44       &  0.40       &  2106  & 18.393 & 0.015        & 18.693 &      0.009   &  1823     & 18.510 & 18.343 &  BS \\ 
CX10 & 16:17:00.412 & -22:58:30.12 &  *          &  *          &  *     & *      & *            & *      &      *       &  *        & *      & *      &  * \\ 
CX11 & 16:17:02.476 & -22:58:39.11 &  0.43       &  0.28       &  1352  & 22.751 & 0.137        & 20.424 &      0.017   &  1134     & 19.980 & 19.278 &  MS \\ 
~    & ~            & ~            &  ~          &  0.33       &  1341  & 22.424 & 0.112        & 20.209 &      0.015   &  1153     & 19.821 & 19.296 &  MS \\ 
~    & ~            & ~            &  ~          &  0.36       &  1283  & 23.065 & 0.158        & 20.560 &      0.033   &  *        & *      & *      &  MS/RG clump \\ 
CX12 & 16:17:02.570 & -22:58:46.25 &  0.43       &  0.10       &  214   & 17.965 & 0.014        & *      &      *       &  *        & *      & *      &  outside HRC \\ 
~    & ~            & ~            &  ~          &  0.24       &  232   & 16.306 & 0.006        & *      &      *       &  *        & *      & *      &  outside HRC \\ 
~    & ~            & ~            &  ~          &  0.33       &  251   & 21.981 & 0.112        & *      &      *       &  *        & *      & *      &  outside HRC \\ 
CX13 & 16:17:01.759 & -22:58:30.54 &  0.42       &  0.11       &  2624  & 23.397 & 0.182        & 22.291 &      0.240   &  *        & *      & *      &  gap \\ 
~    & ~            & ~            &  ~          &  0.25       &  2605  & 24.091 & 0.264        & 22.384 &      0.078   &  *        & *      & *      &  MS/RG clump \\ 
CX14 & 16:17:02.558 & -22:58:31.75 &  0.70       &  0.30       &  2414  & 22.920 & 0.195        & 20.882 &      0.022   &  1871     & 20.555 & 19.802 &  MS \\ 
~    & ~            & ~            &  ~          &  0.30       &  2453  & 22.237 & 0.108        & 19.372 &      0.009   &  1946     & 18.991 & 18.185 &  RG \\ 
~    & ~            & ~            &  ~          &  0.32       &  2452  & 17.661 & 0.011        & 17.380 &      0.004   &  1880     & 16.596 & 16.222 &  BHB \\ 
~    & ~            & ~            &  ~          &  0.38       &  2415  & 22.557 & 0.125        & 20.614 &      0.022   &  *        & *      & *      &  MS/RG clump \\ 
~    & ~            & ~            &  ~          &  0.42       &  2512  & 22.452 & 0.130        & 20.186 &      0.018   &  1978     & 20.135 & 19.291 &  MS \\ 
~    & ~            & ~            &  ~          &  0.64       &  2428  & 22.337 & 0.144        & 20.313 &      0.015   &  1935     & 19.602 & 18.618 &  MS \\ 
~    & ~            & ~            &  ~          &  0.67       &  2541  & 23.253 & 0.346        & 20.673 &      0.038   &  *        & *      & *      &  MS/RG clump \\ 
CX15 & 16:17:02.104 & -22:58:33.05 &  0.43       &  0.15       &  2269  & 23.664 & 0.232        & *      &      *       &  *        & *      & *      &  no NUV \\ 
~    & ~            & ~            &  ~          &  0.17       &  2270  & 22.868 & 0.146        & 20.393 &      0.017   &  *        & *      & *      &  MS/RG clump \\ 
~    & ~            & ~            &  ~          &  0.17       &  2294  & 23.197 & 0.157        & *      &      *       &  *        & *      & *      &  no NUV \\ 
CX16 & 16:17:02.124 & -22:58:21.05 &  0.70       &  0.05       &  3967  & 16.388 & 0.007        & 17.210 &      0.003   &  3338     & 18.350 & 18.285 &  BHB \\ 
~    & ~            & ~            &  ~          &  0.23       &  4786  & 18.383 & 0.020        & 21.117 &      0.048   &  *        & *      & *      &  WD \\ 
CX17 & 16:17:02.224 & -22:58:34.95 &  0.68       &  0.21       &  1944  & 22.971 & 0.235        & 22.555 &      0.158   &  *        & *      & *      &  gap \\ 
~    & ~            & ~            &  ~          &  0.33       &  2005  & 22.324 & 0.157        & 20.989 &      0.057   &  *        & *      & *      &  gap \\ 
~    & ~            & ~            &  ~          &  0.45       &  1952  & 23.262 & 0.221        & 21.275 &      0.045   &  *        & *      & *      &  MS/RG clump \\ 
~    & ~            & ~            &  ~          &  0.47       &  1911  & 23.045 & 0.224        & 20.214 &      0.021   &  *        & *      & *      &  MS/RG clump \\ 
~    & ~            & ~            &  ~          &  0.47       &  2022  & 21.579 & 0.093        & 18.727 &      0.007   &  1975     & 15.046 & 13.441 &  RG \\ 
~    & ~            & ~            &  ~          &  0.51       &  1849  & 16.799 & 0.008        & 18.082 &      0.004   &  *        & *      & *      &  EHB \\ 
~    & ~            & ~            &  ~          &  0.52       &  2050  & 22.752 & 0.209        & 20.713 &      0.026   &  *        & *      & *      &  MS/RG clump \\ 
~    & ~            & ~            &  ~          &  0.53       &  4791  & 22.211 & 0.116        & 20.005 &      0.022   &  *        & *      & *      &  MS/RG clump \\ 
~    & ~            & ~            &  ~          &  0.61       &  1918  & 22.741 & 0.155        & 20.124 &      0.014   &  *        & *      & *      &  MS/RG clump \\ 
CX18 & 16:17:02.824 & -22:58:37.25 &  0.68       &  0.35       &  1559  & 23.374 & 0.221        & 20.390 &      0.027   &  *        & *      & *      &  MS/RG clump \\ 
~    & ~            & ~            &  ~          &  0.38       &  1659  & 23.318 & 0.178        & 20.465 &      0.016   &  1013     & 20.127 & 19.329 &  MS \\ 
~    & ~            & ~            &  ~          &  0.45       &  1601  & 22.161 & 0.090        & 18.628 &      0.005   &  999      & 16.279 & 15.166 &  RG \\ 
~    & ~            & ~            &  ~          &  0.46       &  1531  & 22.672 & 0.122        & 20.249 &      0.020   &  923      & 19.973 & 19.245 &  MS \\ 
~    & ~            & ~            &  ~          &  0.64       &  1607  & 20.351 & 0.040        & 19.881 &      0.012   &  *        & *      & *      &  BS \\ 
CX19 & 16:17:03.854 & -22:58:48.35 &  *          &  *          &  *     & *      & *            & *      &      *       &  *        & *      & *      &  * \\ 
\enddata 
\end{deluxetable}
\end{rotate}

\subsection{CX01: The Nova T\,Sco}

CX01 is particularly interesting since it is located near the site of
nova T\,Sco, which is one of only two novae known to have
occurred along the line of sight to a galactic globular cluster. Based
on an analysis of the historical and HST WFPC2 data, \cite{shara95}
obtained two estimates of the position of the nova, one based on the
offset of the nova from the cluster center, the other based on offsets
from two nearby stars. Based on their estimates of the nova position,
they identified a blue star as the likely post-nova system. Using the
finding chart provided in Shara \& Drissen (1995), we identified this
blue star as source no. 2422 in our $UV$ catalog.

The region containing CX01 and the site of the nova is shown in
Fig.~\ref{fig_tau_sco}. In order to locate the likely position of the
nova in our frames, we offset the Shara \& Drissen (1995) locations
for the nova to our Tycho-based coordinate system using the difference
in the position of the two astrometric reference stars discussed by
Shara \& Drissen (1995). In our coordinate system, the historical nova
position as estimated from the offset to the cluster core is
$16^{h}17{m}02.80^{s}$ $-22^{\circ}58\arcmin32.21\arcsec$ (J2000) and
the position estimated from the two nearby stars is
$16^{h}17^{m}02.84^{s}$ $-22^{\circ}58\arcmin33.21\arcsec$.

Shara \& Drissen's (1995) object is located close to both positions
and is indeed very blue. More specifically, it has $FUV=19.14\pm0.02$
mag and $FUV-NUV=-1.65$ mag, which places it slightly on the blue side
of the WD sequence in Fig.~\ref{cmda}; we have classified it as a hot
WD, a region of the CMD that could indeed contain CVs. However, given
the proximity of the brightest X-ray source, CX01, to the nova
position, it seems highly likely that CX01 is, in fact, associated
with the old nova CV system that produced the 1860 eruption. As shown
in Fig.~\ref{fig_tau_sco}, the position of Shara \& Drissen's (1995)
object is inconsistent with that of CX01. Moreover, Shara \& Drissen
(1995) had already noted that, at M$_B = +6.8$, their source was about
10 times fainter than canonical old novae.

On the other hand, CX01 has a position that is consistent with source
no.~2129, which is one of the ten brightest objects in our $FUV$
catalog, at $FUV = 15.44 \pm0.01$ mag. Furthermore, as shown in
Fig.~\ref{cmdb}, this is the bluest source in our CMD ($FUV-NUV=
-3.80$ mag). This is actually unphysically blue (i.e. significantly
bluer than an infinite temperature blackbody, which has $FUV-NUV=
-1.8$ mag), suggesting that the object must have been much brighter
during the $FUV$ observations than during the $NUV$ observations a
month later. This is surely the counterpart to CX01, and its
variability is consistent with the suggestion by Heinke et al.~(2003)
that this particular source is a CV. The source is close to Shara \&
Drissen's (1995) preferred position for the nova, based on offsets
from nearby stars, and, since this is the type of object that would be
expected to produce a nova, is a much more viable candidate for the
quiescent nova than the blue object identified by Shara \& Drissen
(1995). It is also about 1.5 mag brighter in the $NUV$ than the
candidate described by Shara \& Drissen (1995) and therefore closer to
the quiescent magnitude of other novae. Given its position, X-ray and
UV brightness and variability, this source is almost certainly the
true counterpart to T\,Sco. It clearly merits further study.

\subsection{CX04, CX07, CX13, CX16, CX17: Cataclysmic Variables}
\label{dncx07}

All of the 15 Chandra sources in the $FUV$ field of view have at least
one possible $FUV$ counterpart amongst the $\simeq 3000$ objects in
our $FUV$ catalog if we adopt the 3$\sigma$ X-ray error circle as a
criterion for identifying candidate matches. Thus, it is obvious that
one cannot assume that the identifications suggested in Table
\ref{tab_xray} are real, without considering the magnitude of the
difference in position or the nature of the object. However, CVs are
expected to be found only amongst the gap sources (59 objects) and the
WD candidates (31 objects) in our $FUV$-$NUV$ CMD. Given that three
X-ray sources (CX07, CX13, CX17) are associated with gap sources and a
further two (CX04, CX16) with WD candidates, we regard these
identifications as secure.

The $FUV$ counterpart to CX07 is one of the sources identified with a
gap object (no. 1387 in our catalog, with $FUV= 22.58$ mag). This
$FUV$ object lies only $0\farcs12$ from the best X-ray position. There
is no reason to suspect that the other possible candidate, source
no.~4850, which lies in the RG/MS clump, would have been detected as
an X-ray source. As it happens, however, there is more information in
this case. Source no.~1387 was previously identified by \cite{shara05}
as a CV, which they observed to have undergone a DN outburst. (The
other DN they identified in M\,80 is not within our $FUV$ field of
view.) They suggested that this object, which they called DN1, was
associated with CX17, which is nearby. Our more accurate X-ray
positions make it clear that CX07, and not CX17, is the X-ray source
associated with the dwarf nova.

\subsection{Other X-ray Sources with $FUV$ Counterparts}

There are seven Chandra sources -- CX02, CX06, CX09, CX11, CX14, CX15,
and CX18 -- that are inside both the $FUV$ and $NUV$ fields of view
but have no obvious counterparts. In each case, there is at least one
object within the 3 $\sigma$ error circle but the counterparts are
neither very bright nor in a region of the CMD expected to be
populated by CVs or UV-bright X-ray sources. In several of the cases,
there are no candidates within the 1 $\sigma$ error circle, and this
makes their association with the sources listed in Table
\ref{tab_xray} less likely.

Of these sources, CX02 and CX06 are arguably the most
interesting. Both were identified by \cite{heinke} on the basis of
their hard X-ray spectra and luminosities as possible quiescent
LMXBs. Such objects also are expected to have high values of
$F_{X}/F_{opt}$. The only possible counterparts to these objects are
classified by us as in the MS/RG group, and none of these is within
the 1 $\sigma$ error circle of the X-ray sources. This result is
consistent with the suggestion that they are indeed quiescent LMXBs.

Two further X-ray sources, CX03 and CX12, are located in regions where
there is no $NUV$ coverage. There is a bright $FUV$ object associated
with each of them, but we we cannot classify the object in our
$FUV-NUV$ CMD.

\section{Radial Distributions and Masses of the Stellar Populations}
\label{radial}

\subsection{Radial Distributions}

The radial distribution of the various stellar populations that show
up in our CMD, and also of the X-ray sources, are plotted in
Fig.~\ref{cumulative1}. As our CMD is limited by the $NUV$ data, we
also present the radial distributions for sources brighter than 21.5
mag in the $NUV$. This selection affects only the gap sources. We
compare the BS candidates, the gap sources, and EHB and BHB stars. WD
and MS populations are not shown as both distributions suffer from
incompleteness in the $FUV$ especially in the core region of M\,80 due
to the concentration of bright stars. Such faint sources are not
detectable close to the bright stars because of the broad wings of the
$FUV$ PSF, see Sect.~\ref{data}.

In order to assess the statistical significance of the differences
between the various stellar populations, we applied
Kolmogorov-Smirnov (KS) tests. The KS test measures the probability
that two sample populations are drawn from the same underlying
distribution. Thus the larger the probability, the more similar the
two populations, whereas small probabilities signal significantly
different distributions. The number of sources in the various
distributions in the full and magnitude selected samples are given in
Table~\ref{numbers}; the results from the KS tests are presented in
Table~\ref{KS}.

\begin{deluxetable}{lcc}
\tablecaption{Number of sources in the various populations, both in the
  full samples and the magnitude selected sample with $NUV < 21.5$
  mag. \label{numbers}}
\tablewidth{0pt}
\tablehead{        & all & $NUV < 21.5$}
\startdata
BS      &  75 &  75 \\
BHB     &  80 &  80 \\
EHB     &  30 &  30 \\
EHB1    &  11 &  11 \\
EHB2    &  19 &  19 \\
gap     &  59 &  13 \\
\enddata
\end{deluxetable}

\begin{deluxetable}{lcc}
\tablecaption{KS probability in \% that populations have similar
  radial distributions. \label{KS}} 
\tablewidth{0pt}
\tablehead{ & all    & $NUV < 21.5$}
\startdata
BS--gap    &  0.06 & 95.1  \\
BS--HB     &  0.2  &  0.2  \\
BS--BHB    &  0.02 &  0.02 \\
BS--EHB    & 28.5  & 28.5  \\
BS--CX     & 21.7  & 21.7  \\
gap--HB    & 79.9  &  7.0  \\
gap--BHB   & 77.5  &  4.0  \\
gap--EHB   & 23.4  & 40.2  \\
gap--CX    &  0.9  & 82.8  \\
HB--CX     &  0.3  &  0.3  \\
BHB--EHB   &  9.1  &  9.1  \\
BHB--CX    &  0.2  &  0.2  \\
EHB--CX    &  4.4  &  4.4  \\
EHB1--EHB2 & 49.5  & 49.5  \\
bBS--fBS   &  3.5  &  3.5  \\
\enddata
\end{deluxetable}

Fig.~\ref{cumulative1} shows all of the radial distributions. The BS
stars are clearly the most centrally concentrated population. A strong
concentration of the BSs was already noticed in Ferraro et
al. (1999).

In the magnitude-limited sample (Fig.~\ref{cumulative1}, panel b),
X-ray and gap sources (which include the CV candidates) and BSs are
the most concentrated populations. In fact, the KS test does not
suggest a strong difference between these three populations (see
Table~\ref{KS}). A strong central concentration of BSs and CVs is to
be expected for two reasons. First, a significant fraction of these
objects may be formed in stellar dynamical interactions that
preferentially take place in the dense cluster core. Second, BSs
(merged MS stars) and bright CVs (composed of a WD and near-MS star)
are more massive than ordinary cluster members and will thus sink to
the core due to mass segregation.

No significant differences are found between the radial distributions
of the various HB populations.

\subsection{The Peculiar Blue Straggler Population: Bright versus
  Faint Blue Stragglers}

Bright BSs are thought to be more massive than faint BSs (e.g. Sills
et al. 2000). They are also thought to be younger than the faint BSs,
see e.g. Ferraro et al. (2003, their Fig. 4). Provided that the ages
of all the BSs are larger than the cluster relaxation time
$t_{halfmass}$, the bright, massive BSs should thus be more centrally
concentrated than the fainter, less massive BSs due to mass
segregation. We decided to test this hypothesis.

Since bright BSs are also bluer (and hotter) than faint BSs, we
created our bright BS sample by selecting BSs with $FUV - NUV < 0.9$
mag, and a corresponding faint BS sample by selecting BSs with $FUV -
NUV > 0.9$ mag. Nearly all of the $FUV$ bright and blue BSs are also
optically brighter than $V = 19$ mag (28 out of 33 of the bright and
blue BSs with optical counterparts), and most of the $FUV$ faint and
red BSs are also optically fainter than $V = 19$ mag (11 out of 14
with optical counterparts). 

Fig.~\ref{cumulative1}, panel d, shows the radial distribution of the
bright and faint BSs. Surprisingly, the faint BSs (red line) are more
concentrated than the bright BS (blue line). The KS-test suggests that
there is only a 3.5\% probablility that both the faint and the bright
BSs are drawn from the same parent distribution, see
Table~\ref{KS}. In order to test the sensitivity of this result to the
adopted cluster center, we carried out a Monte Carlo simulation. In
each iteration, we shifted the cluster center randomly in line with
the error derived in Sect.~\ref{centre}, computed the corresponding
new distance from the cluster center for each BS and then calculated
the corresponding KS probability for the faint/red and bright/blue BS
radial distribution. The outcome of 100,000 of these iterations was
that 63\% of the iterations yielded KS probabilities below 4.6\%,
i.e. better than a $2\sigma$ level of confidence. Thus, the marginally
significant difference we find between the radial distributions of
these two types of BSs is not very sensitive to the exact choice of
the cluster center.

This result is rather puzzling. A tentative explanation could be that
BSs get a kick at their formation. This could work, since the bright
BSs are younger and have shorter lifespans than the faint BSs. Thus,
if their initial kick takes BSs out to regions where the relaxation
timescale is shorter than the typical age of faint BSs, but longer
than that of bright BSs, the latter will not have had time to sink
(back) to the core. In any event, we urge others to search for a
similar effect in other globular clusters.

\subsection{Mass Estimates}

The typical mass of objects belonging to a particular stellar
population can be estimated by analyzing their radial
distribution. Assuming that the distributions for all masses can be
approximated by King (1966) models, we can then compare the
distributions of different populations to infer the ratio of their
typical masses. Here, we follow Heinke et al. (2003) and compare our
source distributions to ``generalized theoretical King models'':
 
$S(r) = \int{( 1 + (\frac{r}{r_{c\star}})^{2})^{\frac{1-3q}{2} }} dr$\\

where $q = M_{X}/M_{\star}$, $M_{\star}$ is the mass of the stellar
population that defines the core radius $r_{c\star}$, and $M_{X}$ is
the mass of the source population for which we want to find the mass.

We adopt a core radius of $r_{c\star} = 6\farcs5$ as determined by
Ferraro et al. (1999) from fitting a King model (1996) to their WFPC2
data, and we assume that the core radius is defined by MS turnoff
stars with $M_{\star} = 0.8 M_{\odot}$. In order to have as much
radial coverage as possible, we have corrected the distribution to
account for the fractional area covered by the actual field of view of
the instrument as a function of radius.

The area corrected models are plotted in Fig.~\ref{cumulative2}, with
the radial distributions of the BS, HB, gap and X-ray source
populations overplotted. To avoid confusion, we compare only two
source populations per panel in Fig.~\ref{cumulative2}. BS and HB
distributions are plotted in panel~a. The BS population agrees well
with a model of mass $1.2 M_{\odot}$, while the distribution of HB
stars implies masses around $0.6 M_{\odot}$. Both of these numbers are
reasonable and agree with an average mass estimate based on the mass
distribution along the ZAMS and the ZAHB.

In panel~b, we show the EHB and BHB population. EHB stars seem to be
more massive with $\approx 0.8 M_{\odot}$ than BHB stars ($0.6
M_{\odot}$) (but recall that the difference between these
distributions is not statistically significant). This is contrary to
the mass distribution along the ZAHB, which suggests that the BHB stars
span a mass range of $\approx 0.63 \rm{M}_{\odot}$ to $\approx
0.52\rm{M}_{\odot}$, resulting in an average mass of $\approx 0.58
\rm{M}_{\odot}$. Stars become less massive towards the end of the
ZAHB, so that the EHB stars, on average, should have a mass less than
$\approx 0.51\rm{M}_{\odot}$.

The radial distribution of X-ray sources and of the magnitude-limited
sample of gap sources are plotted in panel~c. Both source populations
seem to be more massive than $1 M_{\odot}$, but a more accurate
estimate is not possible based on our Fig.~\ref{cumulative2}. Our
result broadly agrees with Heinke et al.\ (2003) who suggested an
average mass of $1.2\pm0.2 M_{\odot}$ for the X-ray source population.

The $FUV$ bright vs. $FUV$ faint BSs are compared in panel~d. Based on
this plot, it seems that the faint BSs are more massive ($\approx 1.4
M_{\odot}$) than the bright \& blue BSs ($\approx 1 M_{\odot}$).  A
mass estimate based on the mass distribution along the ZAMS suggests
that the faint and red BSs cover a mass range of 0.94 to 1.13
$\rm{M}_{\odot}$, resulting in an average mass of $\approx 1.04
\rm{M}_{\odot}$, whereas the bright and blue BSs should be more
massive, spanning 1.13 to 1.55 $M_{\odot}$ with an average of $\approx
1.34 \rm{M}_{\odot}$. Again, this is contrary to the mass estimate
based on the radial distribution. However, all of the estimates based
on the radial distributions only hold if the populations have reached
thermal equilibrium. In the ``kick'' scenario sketched in the previous
section (to account for the unexpected difference between bright and
faint BS distributions), the bright BSs do not satisfy this condition.

\section{Conclusions}
\label{summary}

We analyzed deep $FUV$ and $NUV$ images of the core region of
M\,80. We have astrometrically corrected our master images to the
Tycho based WCS, and identified 3168 sources in the $FUV$ master
image, of which 2345 have counterparts in the somewhat smaller $NUV$
master image of M\,80. We have also found optical counterparts for
1268 of the sources in our $FUV - NUV$ CMD. 

The $FUV - NUV$ CMD shows a rich variety of stellar populations in
M\,80. Among the objects are 75 BS candidates, 80 BHB and 30 EHB
stars, 31 WD candidates, and 59 objects in the gap between the WD and
MS. The numbers of bright WDs (24) and of gap sources are consistent
with theoretical predictions. The $FUV - NUV$ CMD reveals clear gaps
along the BHB and EHB (at $T_{eff} \approx 14,500$ K and $\approx
25,500$ K) which can also be identified in the optical CMD.  M\,80
does not appear to have a population of blue hook stars in its core,
as only one possible BHk candidate was found.

Overall, the BS stars are the most centrally concentrated
population, with their radial distribution suggesting a typical blue
straggler mass of about $1.2~M_\odot$. Ferraro et al. (1999, 2003)
suggested that M\,80 comprises an unusual large and concentrated
population of BS stars, compared to other clusters, and suggest that
M\,80 is currently in a transient dynamical state where core collapse
is delayed via stellar interactions that formed the large number of
BSs. We compared our $FUV$ M\,80 data to our $FUV$ M\,15 data, which
covered a similar field of view, and do not find a remarkable excess
in BSs. However, counterintuitively, we found that the faint and red
BSs are significantly more centrally concentrated than the bright and
blue BSs, with only a 3.5\% probablility that faint and bright BSs are
drawn from the same distribution. This result is surprising. One
possible explanation could be that the bright BSs get a kick at their
formation which takes them out to regions where the relaxation
timescale is longer than the typical age of bright BSs but shorter
than the typical age of faint BSs. In that case, the bright BSs would
not have had time to settle towards the cluster core.

Finally, we believe we have recovered the object that was responsible
for the Nova 1860 AD, also known as T\,Scorpii. It is not the UV
bright object identified by \cite{shara95}, but rather a dwarf nova
located at the site of the historical event, which is today the
brightest X-ray object, CX01, in M\,80, see Heinke et
al.\ (2003). This object, source no. 2129, is one of the brightest and
the bluest $FUV$ source in our catalog. This identification has also
enabled us to clearly identify the $FUV$ objects associated with
another 5 of the 15 X-ray sources located in the core of M80. We found
that CX04, CX07, CX13, CX16 and CX17 are associated with gap sources
and WDs. All of these are likely CVs. Our source no. 1387 coincides
with the dwarf nova DN1 observed by \cite{shara05} and is the
counterpart to CX07. For seven X-ray sources (CX02, CX06, CX09, CX11,
CX14, CX15, and CX18) the $FUV$ counterparts are not obvious. The two
remaining X-ray sources, CX03 and CX12 are located in regions where
there is no $NUV$ coverage. There is a bright $FUV$ object associated
with each of them, but we can not classify the object in our $FUV-NUV$
CMD. The radial distributions of the 15 X-ray sources and of the
brighter gap sources ($NUV > 21.5$ mag) are not statistically
different and suggest masses $> 1 M_\odot$.
 
\acknowledgments

This work was supported by NASA through grant GO-10183 from the Space
Telescope Science Institute, which is operated by AURA, Inc., under
NASA contract NAS5-26555. A portion of this work was carried out at
the Kavli Institute for Theoretical Physics in Santa Barbara CA,
USA. This research was supported in part by the National Science
Foundation under Grant No. PHY05-51164.

\pagebreak

\begin{figure}
\centerline{
\includegraphics[width=17cm,height=14.5cm]{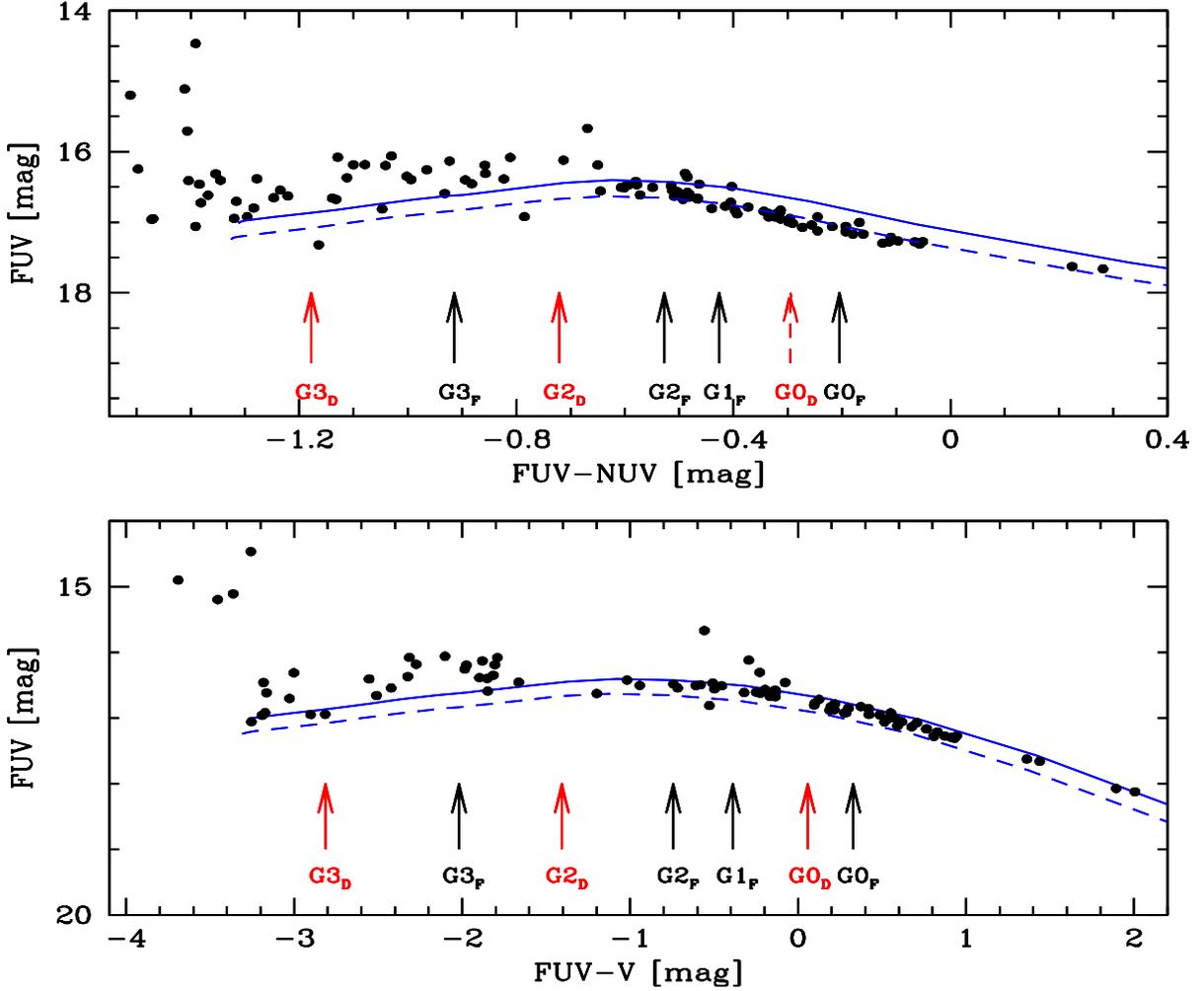}
}
\caption{\label{hb} Top panel: $FUV-NUV$ CMD zoomed in on the HB of
  M\,80. Bottom panel: $FUV - V$ CMD of the HB. In both panels, the
  gaps G$_{\rm{F}}$ suggested by Ferraro et al. (1998) are marked with black
  arrows according to their temperatures. Their location does not
  agree with the gaps G$_{\rm{D}}$, marked with red arrows, in our
  CMDs, but instead are slightly shifted. See the text for details.} 
\end{figure}

\begin{figure}
\includegraphics[width=16cm]{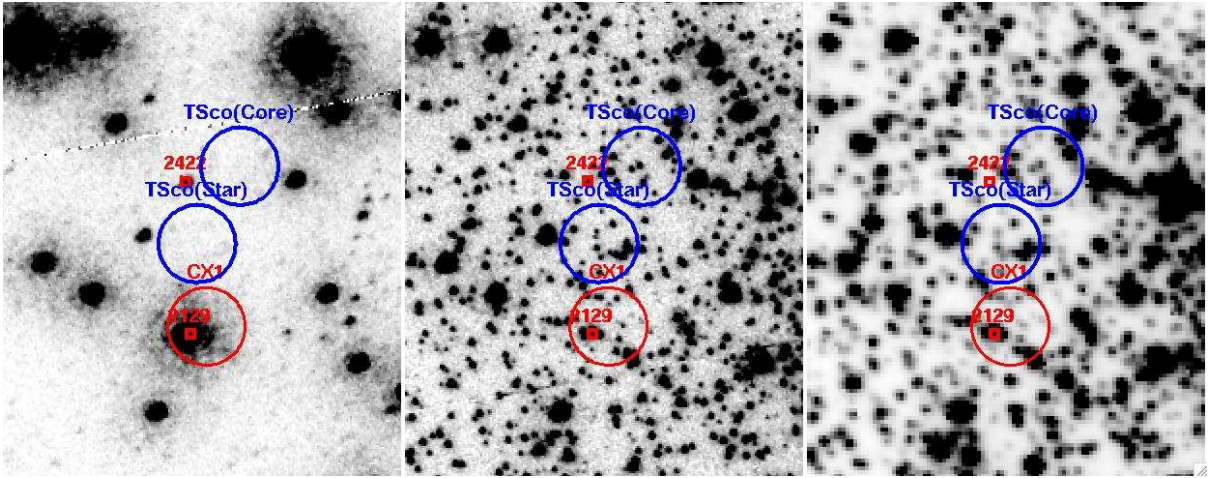}
\caption{\label{fig_tau_sco} The portion of the $FUV$ (left), $NUV$
  (middle) and ACS WFC 435W (right) image where the nova T\,Sco was
  seen. \cite{shara95} suggested that a very blue star, easily visible
  in our $FUV$ and $NUV$ images as object 2422, was likely to be the
  CV responsible for the nova. They also used the historical data to
  make two estimates of the location of the nova within the cluster,
  one based on the location of the cluster core and another based on
  offsets from nearby stars. These are labelled in blue. Our
  astrometry suggests that another very blue object, source no.~2129,
  is more likely to be the brightest X-ray source in the globular
  cluster, marked with a red circle. Given the uncertainties in
  locating the nova from the historical record, our object 2129 seems
  a better candidate as the CV which gave rise to the nova.}
\end{figure}

\begin{figure}
\centerline{
\includegraphics[width=16cm]{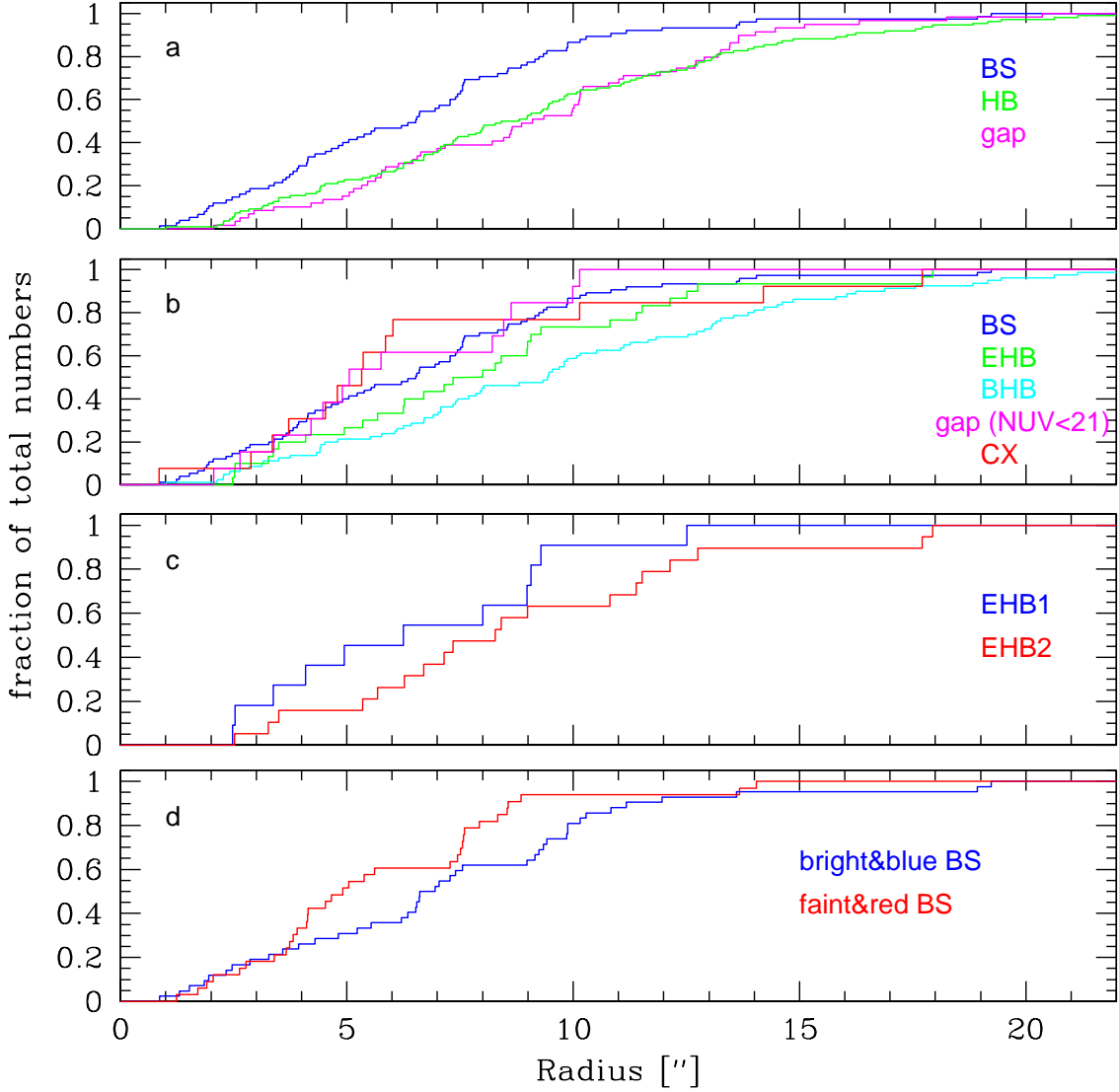}
}
\caption{\label{cumulative1} Cumulative radial distributions of the
  various stellar populations that show up in our $FUV-NUV$ CMD, and
  of the X-ray sources. We only compare the BS candidates, gap
  objects, HB stars and X-ray sources. Panel (a) shows the radial
  distribution of BSs, HB stars and gap sources. Panel (b) shows the BHB
  and EHB populations, the magnitude selected ($NUV < 21$ mag) gap
  sources, the BSs and X-ray sources. Panel (c) compares the EHB1 and
  EHB2 populations. Panel (d) compares the radial distribution of the
  bright/blue BSs and the faint/red BSs. Contrary to our expectation,
  we see the faint (red) BSs to be stronger concentrated than the
  bright (blue) BSs. See the text for the details.} 
 \end{figure}

\begin{figure}
\centerline{
\includegraphics[width=16cm]{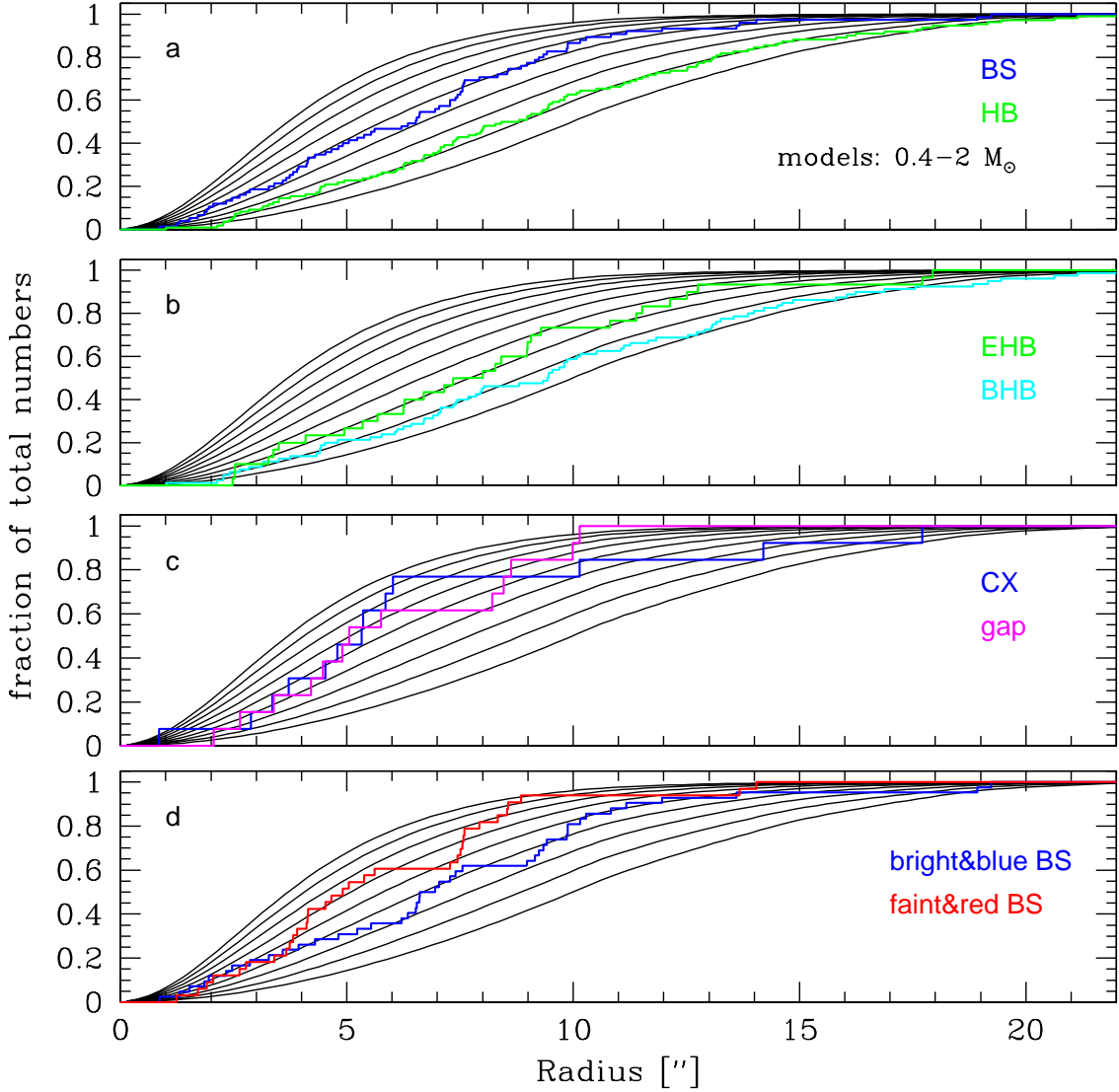}
}
\caption{\label{cumulative2} Comparison of the source distribution
  with theoretical King models for stellar populations with average
  masses ranging from $0.4 M_{\odot}$ (bottom black line in each
  panel) to $2 M_{\odot}$ (top black line), in steps of $0.2
  M_{\odot}$. BS and HB populations are plotted in panel~(a). As can be
  seen, the BS population agrees well with a model of mass $1.2
  M_{\odot}$, and the HB stars with masses around $0.6
  M_{\odot}$. Panel~(b) shows the EHB and BHB population. EHB stars seem
  to be slightly more massive ($0.8 M_{\odot}$) than BHB stars ($0.6
  M_{\odot}$). Panel~(c) shows the X-ray and the magnitude selected gap
  sources. Both source populations seem to be more massive than $1
  M_{\odot}$. Panel~(d) shows the bright vs. faint BSs. Surprisingly,
  the faint BSs seem to be more massive ($\approx 1.4 M_{\odot}$) than
  the bright BSs ($\approx 1 M_{\odot}$). See the text for details.}
\end{figure}

\end{document}